\documentclass[twocolumn,longbibliography]{revtex4-2}

\usepackage{hyperref}
\usepackage{graphicx}
\usepackage{epsf}
\usepackage{amsmath,amssymb}
\usepackage{color}
\newcommand{\bvec}{\boldsymbol}

\def\Pb#1#2#3{{}^{#1#2#3}\textrm{Pb}}
\def\Ni#1#2{{}^{#1#2}\textrm{Ni}}
\def\Ca#1#2{{}^{#1#2}\textrm{Ca}}
\def\O#1#2{{}^{#1#2}\textrm{O}}

\def\Sn#1#2#3{{}^{#1#2#3}\textrm{Sn}}
\def\Sr#1#2{{}^{#1#2}\textrm{Sr}}
\def\Zr#1#2{{}^{#1#2}\textrm{Zr}}

\def\Ni#1#2{{}^{#1#2}\textrm{Ni}}

\usepackage{ulem}

\begin{document}
\title{Effects of density-dependent spin--orbit interactions 
in Skyrme--Hartree--Fock--Bogoliubov calculations 
of the charge radii and densities of Pb isotopes}

\author{Yoshiko Kanada-En'yo}
\affiliation{Department of Physics, Kyoto University, Kyoto 606-8502, Japan}

\begin{abstract}

I have investigated the $N$-dependence of the charge radii $r_c$ across $N=126$ in the Pb isotopes and
the surface densities of $\Pb208$ using
new Skyrme interactions that contain a density-dependent spin--orbit term, which I have named the
Skyrme--ddso interactions.
I have compared the results obtained using Skyrme--Hartree--Fock--Bogoliubov calculations
that employ the Skyrme-ddso interactions with the original Skyrme results and have discussed
the effects of including the density-dependent spin--orbit term.
The results for the kink behavior of $r_c$ at $N=126$ in the Pb isotopes were improved
by the inclusion of the density-dependent spin--orbit term.
Moreover, the new Skyrme calculations yield
better results for the inner part of the surface proton density of $\Pb208$ at $r\sim 5$~fm.
The change in the $\ell s$ potentials from the original Skyrme calculation
contributes to the kink behavior through its effects on the $\ell s$ splitting of the neutron $2g_{7/2,9/2}$ orbitals.
It also affects the inner part of the surface proton density through its effect on the proton $1h_{11/2}$ orbitals.
I have further demonstrated that the change in the isoscalar-to-isovector ratio of the spin--orbit term
makes only a minor contribution to the single-particle energies
and to the kink behavior in the Pb isotopes.
In addition, I have investigated $\Ca48$ using Skyrme--Hartree--Fock calculations with the new
Skyrme interactions and have determined the effects of the density-dependent spin--orbit term on its radii and densities.
\end{abstract}

\maketitle

\section{Introduction}

Ground-state properties such as the binding energy and root-mean-square (rms)
radii of the nucleon distributions are fundamental observables of nuclei, and they have been 
studied extensively
over a wide region of the nuclear chart, including unstable nuclei.
Experimental data for those properties
have been utilized to improve energy-density functionals for both non-relativistic and
relativistic mean-field approaches.
The rms radii of neutrons and protons in nuclei with
$N>Z$ have been attracting particular interest, especially for obtaining
information such as the symmetry energy of nuclear matter.
Intensive studies of the nuclear radii of doubly magic nuclei---including $\Pb208$, $\Sn132$, and $\Ca48$---
have been performed using various theoretical approaches.
In recent years, there have been many experimental attempts to determine
the rms radii ($r_n$) and density distributions ($\rho_n$) of the neutrons in $\Pb208$ and $\Ca48$
using hadronic and electronic probes. However, the extracted values still contain
large uncertainties because of model ambiguities in the analyses as well as statistical errors.
In contrast, the rms radii ($r_p$) and densities ($\rho_p$) of protons
have been determined precisely from electron-scattering experiments. Moreover,
$x$-ray measurements have been widely
used to obtain high-quality data for the isotope shifts of the charge radii.
Using a laser-cooling technique, charge radii have been 
measured recently for neutron-rich nuclei, including Sn isotopes across $N=82$~\cite{Gorges:2019wzy} and Ca isotopes across
$N=28$~\cite{GarciaRuiz:2016ohj}.

The $N$-dependence of the charge radii across
$N=126$ is a well-known and long-discussed problem; the experimental data revealed the existence of a kink behavior in the charge radii of the Pb isotopes at $N=126$ ($\Pb208$).
Although relativistic mean-field~(RMF) calculations have succeeded in describing this kink phenomenon, 
many Hartree--Fock--Bogoliubov~(HFB) calculations using conventional Skyrme interactions have failed.
Reinhard {\it et al.} investigated this problem by comparing
the non-relativistic and relativistic mean-field calculations~\cite{Reinhard:1995zz},
and they demonstrated that the small energy difference between the neutron $1h_{11/2}$ and $2g_{9/2}$ orbitals
in $\Pb208$ is essential for reproducing the kink behavior of the charge radii.
In the usual non-relativistic approaches,
the neutron spin--orbit potential is proportional to $\nabla \rho+\nabla \rho_n$, as derived from 
zero-range spin--orbit nucleon--nucleon ($NN$) interactions. On the other hand, in the relativistic calculations,
the isoscalar~(IS) component---which corresponds to $\nabla \rho$ in the Skyrme formalism---was found to
be dominant.
To modify the ratio of isoscalar-to-isovector (IS/IV) components of the spin--orbit potential
in the Skyrme calculations,
the investigators proposed a new Skyrme energy-density functional (EDF), with the extended form
$b_4\nabla \rho+b^\prime_4\nabla \rho_q$ ($q=n,p$) for the spin--orbit term, instead of the
usual form ($b_4=b^\prime_4$) in the conventional Skyrme parametrization. Using interactions
such as the
SkI3 interaction with $b^\prime_4=0$ for the IS-type and the SkI4 interaction
with $b_4\sim -b^\prime_4$ for the reverse-type,
they obtained better results for the kink phenomenon in the Pb isotopes.
The latter interaction (SkI4) yielded results similar to the RMF results for the single-particle energies (SPEs)
of $\Pb208$, but $b_4\sim -b^\prime_4$ seems an unrealistic choice.
The role of spin--orbit potentials and neutron SPEs in the kink phenomenon in Pb isotopes
also has been discussed by Goddard {\it et al.}~\cite{Goddard:2012dk}. 
They showed that a modified version 
of the SLy4 interaction that uses an IS-type spin--orbit term~\cite{Chabanat:1997un}---called SLy4mod--- reproduced the kink in the charge radii.
However, in that case, it was not necessary to
change the IS/IV ratio of the spin--orbit term, but the strength of the spin--orbit interactions must be reduced to
reproduce the kink; the SLy4mod interaction adopts a 20\% weaker strength for the spin--orbit term
than that used in the original SLy4 interaction.

A similar kink behavior of the charge radii has been observed
for the Sn isotopes at $N=82$, which 
remains an open problem.
To describe the $N$-dependence of the charge radii in the Sn isotopes,
new EDFs have been developed that go beyond the conventional relativistic and non-relativistic models; for instance,
by including the $\delta$ meson in the meson-exchange model
in the relativistic approach~\cite{Reinhard:2017ugx,Perera:2021ztx} and
by adopting the Fayans EDF in the non-relativistic approach \cite{Reinhard:2017ugx,Gorges:2019wzy,Kortelainen:2021raz}.

Nakada {\it et al.} employed a density-dependent
spin--orbit interaction in HFB calculations with finite-range effective $NN$ interactions,
and they obtained improved results for the isotope shift of the charge radii ~\cite{Nakada:2014apa,Nakada:2015cla}.
They adjusted the density dependence of the spin--orbit interaction 
phenomenologically to reproduce the observed $\ell s$-splitting of the neutron $1h_{11/2,13/2}$ orbitals in $\Pb208$,
although they justified the density dependence in terms of a contribution from three-nucleon spin--orbit forces
to the effective $NN$ spin--orbit interactions \cite{Kohno:2012vj}.

High-precision data for the charge densities contain
further information beyond just the charge radii,
which can be utilized to test the EDFs,
as argued in Ref.~\cite{Reinhard:2019ixi}. 
Yoshida {\it et al.} investigated
detailed profiles of the surface proton and neutron densities
of $\Pb208$ and $\Ca48$ and tested the EDFs used in the
relativistic and non-relativistic approaches.
Compared with the experimental data for proton densities determined from electron scattering,
they found that the Skyrme interactions tend to underestimate the inner region of the surface proton densities
of $\Pb208$ at $r\sim 5$~fm and of $\Ca48$ at $r\sim 2.5$~fm.
On the other hand, RMF calculations using interactions such as DD--ME2~\cite{Lalazissis:2005de}
and NL3~\cite{Lalazissis:1996rd} obtained better agreement with the experimental data for the inner parts of the surface
proton densities.

Thus, different trends have been 
found in the results for the charge radii and surface densities
between relativistic and non-relativistic EDFs;
the Skyrme EDFs often underestimate the kink behavior of the charge radii and the inner-surface proton densities,
while relativistic EDFs tend to obtain better results.
Spin--orbit interactions, which are treated in quite different ways
in the relativistic and non-relativistic frameworks, are considered to be a major source of these differences.

In the present paper, I introduce a density-dependent strength for the spin--orbit term in the Skyrme EDFs
and examine how the charge radii and densities are affected by the inclusion of this term.
I first reexamine the standard Skyrme--HFB (SHFB) results for the Pb isotopes in comparison with the RMF results.
I discuss the results obtained for $\Pb208$ from these
SHFB calculations using the
SLy4~\cite{Chabanat:1997un} and SkM*~\cite{Bartel:1982ed} interactions,
which are widely used conventional parametrizations,
and also using other Skyrme parametrizations, including the SAMi~\cite{Roca-Maza:2012dhp}, SkO$^\prime$~\cite{Reinhard:1999ut}, SKRA~\cite{Rashdan:2000}, SkI series~\cite{Reinhard:1995zz}, SkT series~\cite{Tondeur:1984gwk}, and Skxs20~\cite{Brown:2007zzc}
interactions.
For comparison, I also discuss the results obtained from relativistic Hartree--Bogoliubov~(RHB) calculations
using the DD--ME2~\cite{Lalazissis:2005de} and DD--PC1~\cite{Niksic:2008vp} interactions.
To compare the single-particle potentials between various approaches---including the non-relativistic
framework with zero-range interactions and the relativistic framework
with finite-range interactions for the meson-exchange model---
I calculate the effective single-particle potentials defined by the single-particle densities.
I then propose a modification of the original Skyrme interactions
that incorporates a spin--orbit term with a density-dependent strength into the standard Skyrme EDF,
and I investigate how the SPEs and charge densities of $\Pb208$ are affected by the inclusion of this
density-dependent spin--orbit term.
I also performed SHFB calculations, both with and without the density-dependent spin--orbit term, and
RHB calculations for the Pb isotopes to discuss the kink behavior of the charge radii at $N=126$.
In addition, I have investigated the charge density and radius of $\Ca48$ using the new Skyrme EDFs with
the density-dependent spin--orbit term.

This paper is organized as follows. The methods of calculation
are explained in Section~\ref{sec:calculations}.
In Section~\ref{sec:results}, the results obtained
from SHFB calculations using the original Skyrme interactions
are presented and compared with the RHB results.
In Section~\ref{sec:new-results}, the results obtained using the new Skyrme EDFs
with the density-dependent spin--orbit term are presented and compared with the original results.
A summary is provided in Section~\ref{sec:summary}.
Appendix \ref{sec:parameters} provides the parametrization used in the present SHFB calculations.

\section{Methods of calculation}\label{sec:calculations}

\subsection{Computational codes for HFB and RHB calculations}

I performed spherical SHFB and RHB calculations for even--even nuclei 
using the computational codes HFBRAD~\cite{Bennaceur:2005mx}
and DIRHB~\cite{Niksic:2014dra}, respectively.
To perform the SHFB calculations with the new interactions,
I added the density-dependent spin--orbit terms to the HFBRAD code.

\subsection{EDF for SHFB calculations}
In the SHFB formalism, energy densities
are expressed in terms of the local particle densities $\rho(\bvec r)$
and pairing densities $\tilde{\rho}(\bvec r)$;
the normal and abnormal kinetic-energy densities $\tau(\bvec r)$ and $\tilde{\tau}(\bvec r)$;
and the spin-current vectors $\bvec J(\bvec r)$ and $\tilde{\bvec J}(\bvec r)$.
Detailed definitions are given in Ref.~\cite{Bennaceur:2005mx}.

The kinetic-energy density is given by
\begin{align}
{\cal K}=\frac{\hbar^2}{2m}\tau-\frac{\hbar^2}{2mA}\tau,
\end{align}
where $m$ is the nucleon mass.
The second term is the center-of-mass (c.m.) kinetic-energy
correction, which is taken into account before performing the energy variation in simple treatments as done in 
such cases that use the SLy4 and SkM* interactions. However, other treatments
of the variation---without the c.m, correction---
are adopted in some cases, such as in the SkI series.
Because the major interests in the present paper are the nucleon-density distributions and nuclear radii,
I controlled the selection of parameters in the HFBRAD code to perform
the energy variation both with and without the c.m. kinetic-energy correction in the SHFB calculations.

The conventional Skyrme energy density ${\cal E}_\textrm{Skyrme}$ utilizes
the Skyrme parameters $t_0$, $t_1$, $t_2$, $t_3$, $x_0$, $x_1$, $x_2$, $x_3$, $\gamma$, and $W_0$,
as given in Eq.~(19) of Ref.~\cite{Bennaceur:2005mx}.
The spin--orbit term in ${\cal E}_\textrm{Skyrme}$ is given by
\begin{align}\label{eq:edf_so3o}
{\cal E}^\textrm{so}_\textrm{Skyrme}=\frac{1}{2} W_0\left(
\bvec{J}\cdot\nabla\rho +\sum_q \bvec{J}_q\cdot\nabla\rho_q\right),
\end{align}
where the index $q$ represents either neutrons or protons, while densities without this index represent the
total (isoscalar) densities.
To modify the IS/IV ratio of the spin--orbit term, I employ
another expression that uses
the parameters $b_4$ and $b^\prime_4$ instead of $W_0$: 
\begin{align}\label{eq:edf_so}
{\cal E}^\textrm{so}_\textrm{Skyrme}=b_4
\bvec{J}\cdot\nabla\rho +b_4^\prime \sum_q \bvec{J}_q\cdot\nabla\rho_q.
\end{align}
The choice $b_4 = b^\prime_4 =W_0/2$ is equivalent to the usual spin--orbit term
of the conventional Skyrme parametrization,
whereas $b^\prime_4=0$ is employed for the IS-type spin-orbit term.

The SHFB calculations employed
surface-type pairing forces. For the
SLy4 and SkM* interactions, I used the default pairing interactions
in the HFBRAD code~\cite{Bennaceur:2005mx}.
For other Skyrme interactions, I adopted the same surface-type pairing interactions
as those in SLy4 but multiplied by a factor $\theta_\textrm{pairing}$, which I used to adjust the mean pairing gap of the neutrons in $\Sn120$.
In the present paper, I performed
the SHFB calculations for the Pb isotopes using the adjusted pairing forces,
while I performed the SHF calculations 
for the doubly closed Ca isotopes $\Ca40$, and $\Ca48$ without the pairing forces.

The parameter sets for the SHFB calculations adopted in the present paper are listed in Tables~\ref{tab:skyrme-int1}
and \ref{tab:skyrme-int2} of Appendix \ref{sec:parameters}.
I employed the Skyrme interactions
SLy4, SkM*, SkI2, SkI3, SkI4, SAMi, SkO$^\prime$,
SKRA, Skxs20, SkT1 (T1), SkT2 (T2), SkT3 (T3), SkT4 (T4), SkT5 (T5), and SkT6 (T6).
I also used a SLy4-like version---named the SLy4--IS interaction---in which the usual spin--orbit term
of the SLy4 interaction was replaced with an IS-type term with $b_4=93$~MeV and $b^\prime_4=0$.
I used the strength of this term in the SLy4--IS interaction to
adjust the spin--orbit component of the total energy of $\Pb208$ to the value obtained using the original SLy4 interaction.
The value $b_4=93$~MeV is approximately equal to $b_4=3W_0/4=92.5$~MeV derived from
the ansatz $\rho_p=(Z/N)\rho_n$.
For reference, I also tested the interaction SLy4mod,
which is another SLy4-like version that employs the IS-type spin--orbit term
used in Ref.~\cite{Goddard:2012dk}.
The difference between the SLy4--IS and SLy4mod interactions involves only the
strength $b_4$ of the IS-type spin--orbit term;
the latter employs the 20\% smaller value $b_4=75$~MeV,
which means that the spin--orbit interaction is weaker in the SLy4mod interaction than in the original SLy4 interaction.

\subsection{New Skyrme EDF with a density-dependent spin--orbit term}

I incorporated the density-dependent strength of the spin--orbit term into
the Skyrme energy density in a way similar to that employed in Ref.~\cite{Nakada:2014apa}:
\begin{align}
&{\cal E}^{\textrm{so}(\rho)}_\textrm{Skyrme}
=\frac{1}{4}
D(\rho) \Bigl[
x_4 (\rho\nabla\cdot \bvec{J}-\bvec{J}\cdot\nabla\rho)
\nonumber\\
&+ (1-x_4) \sum_q(\rho_q\nabla\cdot \bvec{J}_q-\bvec{J}_q\cdot\nabla\rho_q)\Bigr],
\end{align}
where $D(\rho)$ represents the density-dependent strength of the interaction, and $x_4$ is a parameter that changes
the IS/IV ratio.
For $x_4=0.5$ and $D(\rho)= {\rm constant} =-W_0$, this equation
reduces to Eq.~\eqref{eq:edf_so3o} for the conventional Skyrme interaction, which contains the usual spin--orbit term.
In the present paper, I employed $x_4=1$ to give the IS-type ratio of the
density-dependent part as
\begin{align}
&{\cal E}^{\textrm{so}(\rho)}_\textrm{Skyrme}
=\frac{1}{4}
D(\rho) \Bigl[
(\rho\nabla\cdot \bvec{J}-\bvec{J}\cdot\nabla\rho)\Bigr],
\end{align}
and I rewrote the spin--orbit energy
$E^{\textrm{so}(\rho)}_\textrm{Skyrme} =\int d \bvec{r}
{\cal E}^{\textrm{so}(\rho)}_\textrm{Skyrme}$
as
\begin{align}
&E^{\textrm{so}(\rho)}_\textrm{Skyrme}
= - \frac{1}{2}\int d\bvec{r}
{\cal D}(\rho)
\bvec{J}\cdot\nabla\rho,\\
&{\cal D}(\rho) \equiv D(\rho)+\frac{1}{2}\frac{\delta D(\rho)}{\delta \rho} \rho.
\end{align}
For spherical nuclei, the density-dependent spin--orbit term contributes
to the single-particle potentials, which contain both central and spin-obit parts:
\begin{align}
&U^{\textrm{so}(\rho)}(r) =
U^{\textrm{so}(\rho)}_\textrm{cent}(r)+U^{\textrm{so}(\rho)}_{\ell s} (r)
\bvec{\ell}\cdot\bvec{s}, \\
& U^{\textrm{so}(\rho)}_\textrm{cent} (r) = \frac{{\cal D}(\rho) }{2}
\nabla\cdot \bvec{J}, \\
&U^{\textrm{so}(\rho)}_{\ell s} (r) = - \frac{1}{r}\frac{{\cal D}(\rho)}{2}
\frac{d\rho}{dr}.
\end{align}
In the present paper, I adopted $D(\rho)=-w_4\rho^{\gamma_4}$ corresponding to
\begin{align}
{\cal D}(\rho)=-w_4\left( 1+\frac{\gamma_4}{2} \right)\rho^{\gamma_4}.
\end{align}

To examine the effects of the density--dependent spin-orbit term,
I considered the new Skyrme EDFs obtained
by changing the spin--orbit term ${\cal E}^\textrm{so}_\textrm{Skyrme}$
of the original Skyrme parametrization
as follows:
\begin{align}\label{eq:edf_ddso-a}
&{\cal E}^\textrm{so}_\textrm{Skyrme}\to {\cal E}^{\textrm{ddso}}_\textrm{Skyrme}=0.33 {\cal E}^\textrm{so}_\textrm{Skyrme}
+{\cal E}^{\textrm{so}(\rho)}_\textrm{Skyrme},\\
&{\cal E}^{\textrm{so}(\rho)}_\textrm{Skyrme}
\equiv -\frac{1}{2}
{\cal D(\rho)} \bvec{J}\cdot \nabla \rho,
\end{align}
in which the original spin--orbit term is reduced to 33\% and the density-dependent term is appended.
I label this new version containing the density-dependent spin--orbit term with the IS-type ratio
``Skyrme--ddso.''
Similarly, the new interaction constructed from the SLy4 parametrization is called SLy4--ddso.
I adjusted the strength $w_4$ of the density-dependent part of this interaction
to obtain the same value of the spin--orbit energy
for $\Pb208$ as that obtained using the original Skyrme interaction.
I also tested an additional version, labeled ``Skyrme--ddso2,'' which has the usual ratio~($x_4=0.5$)
of the density-dependent spin--orbit term:
\begin{align}\label{eq:edf_ddso-b}
&{\cal E}^{\textrm{ddso2}}_\textrm{Skyrme}=0.33 {\cal E}^\textrm{so}_\textrm{Skyrme}\nonumber\\
&-
\frac{1}{2}{\cal D(\rho)} \Bigl( 0.5 \bvec{J}\cdot \nabla \rho + 0.5 \sum_q \bvec{J}_q\cdot \nabla \rho_q\Bigr ).
\end{align}
The values $w_4$ and $x_4$ in the Skyrme--ddso and Skyrme--ddso2 interactions are listed in
Tables~\ref{tab:skyrme-int1} and \ref{tab:skyrme-int2}.
Because the Skyrme--ddso and
Skyrme--ddso2 interactions yield results that are qualitatively similar to each other, I mainly present
the Skyrme--ddso results in this paper.

\subsection{Effective single-particle potentials} \label{subsec:eff-pot}

It is usually not trivial to compare
single-particle potentials between different approaches 
of relativistic and non-relativistic frameworks using 
finite-range and zero-range nuclear interactions, respectively.
To discuss the single-particle potentials in the
SHFB and RHB approaches on an equal footing,
I therefore defined the following effective single-particle potentials for the single-particle
orbitals:
\begin{align}
&U^\textrm{eff}_{\alpha}(r)=\epsilon_\alpha-\frac{\hbar^2}{2m}\Bigl[
-\frac{1}{u_\alpha(r)}\frac{d^2}{dr^2}u_\alpha(r)+\frac{\ell(\ell+1)}{r^2}\Bigr], \\
&u_\alpha(r)\equiv r\sqrt{\rho^\textrm{sp}_{\alpha}(r)},
\end{align}
where $\epsilon_\alpha$ and $\rho^\textrm{sp}_\alpha$ are the
single-particle energy and the density of the orbital labeled $\alpha$, which I
obtained from the SHFB and RHB calculations.
For the proton orbitals, the Coulomb potential part is subtracted.
The effective potential $U^\textrm{eff}_{\alpha}(r)$
provides the equivalent
single-particle energy $\epsilon_\alpha$ and density $\rho^\textrm{sp}_\alpha(\bvec{r})=|\psi_\alpha(\bvec{r})|^2$
in a single-particle potential model,
\begin{align}
\epsilon_\alpha\psi_\alpha(\bvec{r})=-\frac{\hbar^2}{2m}\nabla^2 \psi_\alpha(\bvec{r}) +U^\textrm{eff}_{\alpha}\psi_\alpha(\bvec{r}).
\end{align}

The effective single-particle potentials so defined are orbital-dependent, and they contain
finite-range (or $k^2$-dependent) interaction effects
in addition to the mean-field potentials. In the SHFB framework,
$U^\textrm{eff}$ can be written as
\begin{align}
&U^\textrm{eff}_{\alpha}=U^\textrm{MF}+\Bigl(\frac{\hbar^2}{2m^*}-\frac{\hbar^2}{2m}\Bigr)\frac{\tau^\textrm{sp}_\alpha}{\rho^\textrm{sp}_\alpha},\\
&\tau^\textrm{sp}_\alpha=-\psi^*_\alpha \nabla^2 \psi_\alpha,
\end{align}
where $U^\textrm{MF}$ is the so-called mean-field~(HF) potential
that appears in the HF equation,
$\tau^\textrm{sp}_\alpha$ represents the single-particle kinetic-energy density,
and $m^*$ is the effective mass defined by
\begin{align}
&\frac{\hbar^2}{2m^*}=\frac{\hbar^2}{2m} +\frac{t_1}{4}\Bigl[ \bigl(1+\frac{x_1}{2} \bigr)\rho
-\bigl(x_1+\frac{1}{2} \bigr)\rho_q \Bigr]\nonumber\\
&+\frac{t_2}{4}\Bigl[ \bigl(1+\frac{x_2}{2} \bigr)\rho
+\bigl(x_1+\frac{1}{2} \bigr)\rho_q \Bigr].
\end{align}

In the relativistic framework, the effective potentials $U^\textrm{eff}_{\alpha}(r)$ also contain
the coupled-channel effects of the large [$f_\alpha(r)$] and small [$g_\alpha(r)$] components of the Dirac spinors. In addition, there is an ambiguity in the definition of $u_\alpha(r)$.
In the present prescription, I chose
$u_\alpha(r)=\sqrt{f^2_\alpha(r)+g^2_\alpha(r)}$ using the single-particle baryon density
$\rho^\textrm{sp}_{\alpha}(r)=f^2_\alpha(r)+g^2_\alpha(r)$.
An alternative choice is $u_\alpha(r)=f^2_\alpha(r)/r$,
but these two expressions for $u_\alpha(r)$
produce only minor differences in the resulting $U^\textrm{eff}_{\alpha}(r)$.

I evaluated the effective $\ell s$ potential $U^\textrm{eff}_{\ell s}$ from the
potential difference between the $\alpha_>=n\ell_{j_>}$ $(j_>=\ell+1/2)$ orbital
and the $\alpha_<=n\ell_{j_<}$ $(j_>=\ell-1/2)$ orbital:
\begin{align}
U^\textrm{eff}_{\alpha,\ell s}=\frac{2}{2\ell+1}\Bigl[U^\textrm{eff}_{\alpha_>}(r)-U^\textrm{eff}_{\alpha_<}(r)
\Bigr],
\end{align}
which represents the $r$-dependent part of the spin--orbit potential.
The effective potential $U^\textrm{eff}_\alpha(r)$ can then be expressed as the
sum of a central and a spin--orbit part:
\begin{align}
&U^\textrm{eff}_{\alpha}(r)=U^\textrm{eff}_{\alpha,\textrm{av}}(r)+U^\textrm{eff}_{\alpha,\ell s}(r)
\bvec{l}\cdot\bvec{s}.
\end{align}
Here, $U^\textrm{eff}_{\alpha,\textrm{av}}(r)$ is the averaged potential of the $\alpha_>$ and $\alpha_<$
orbitals given as
\begin{align}
U^\textrm{eff}_{\alpha,\textrm{av}}=\frac{\ell+1}{2\ell+1}U^\textrm{eff}_{\alpha_>}
-\frac{\ell}{2\ell+1}U^\textrm{eff}_{\alpha_<}.
\end{align}

\section{Results from the original Skyrme--HFB and RHB calculations} \label{sec:results}

\subsection{Density distributions and potentials}

\begin{figure}[!htpb]
\includegraphics[width=8.6 cm]{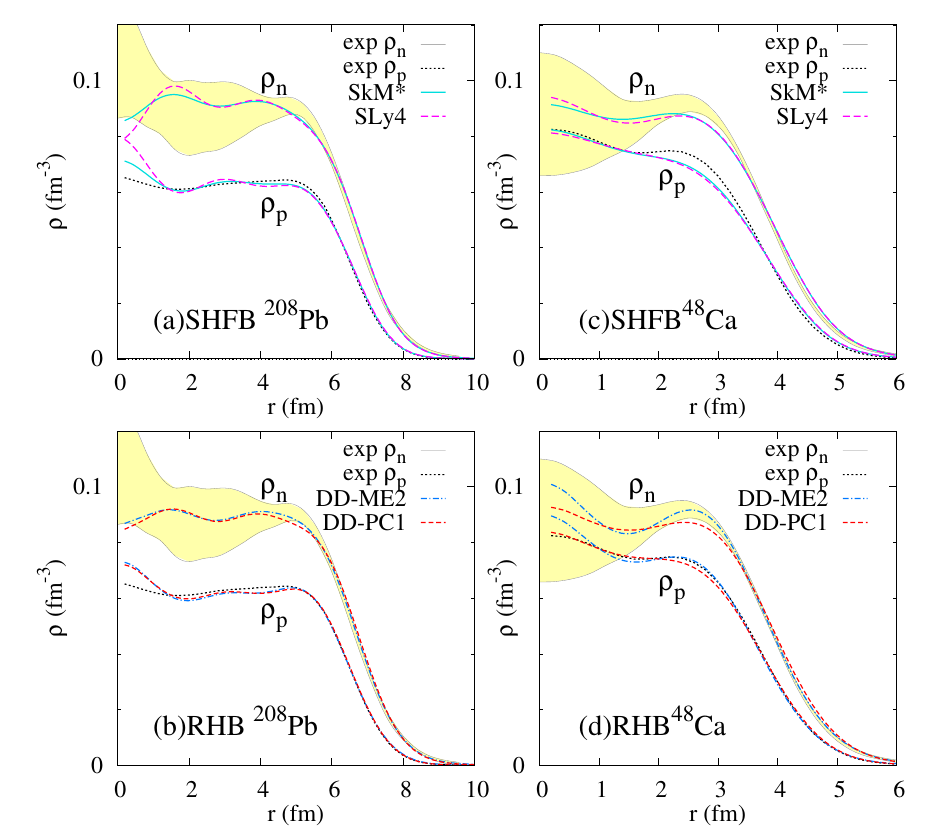}
\caption{
Neutron [$\rho_n(r)$] and proton [$\rho_p(r)$] densities
of (a) $\Pb208$ and (c) $\Ca48$ obtained from SHFB
calculations using the SLy4 and SkM* interactions, and those
of (b) $\Pb208$ and (d) $\Ca48$
obtained from
RHB calculations using the DD--ME2 and DD--PC1 interactions.
The experimental proton densities are those determined
from electron-scattering data \cite{DeJager:1987qc}, and
the neutron densities (with error envelopes) are those
extracted from $(p,p)$ data at $E_p=295$~MeV~\cite{Zenihiro:2010zz,Zenihiro:2018rmz}.
\label{fig:dens-pn}}
\end{figure}

\begin{figure}[!htpb]
\includegraphics[width=8.6 cm]{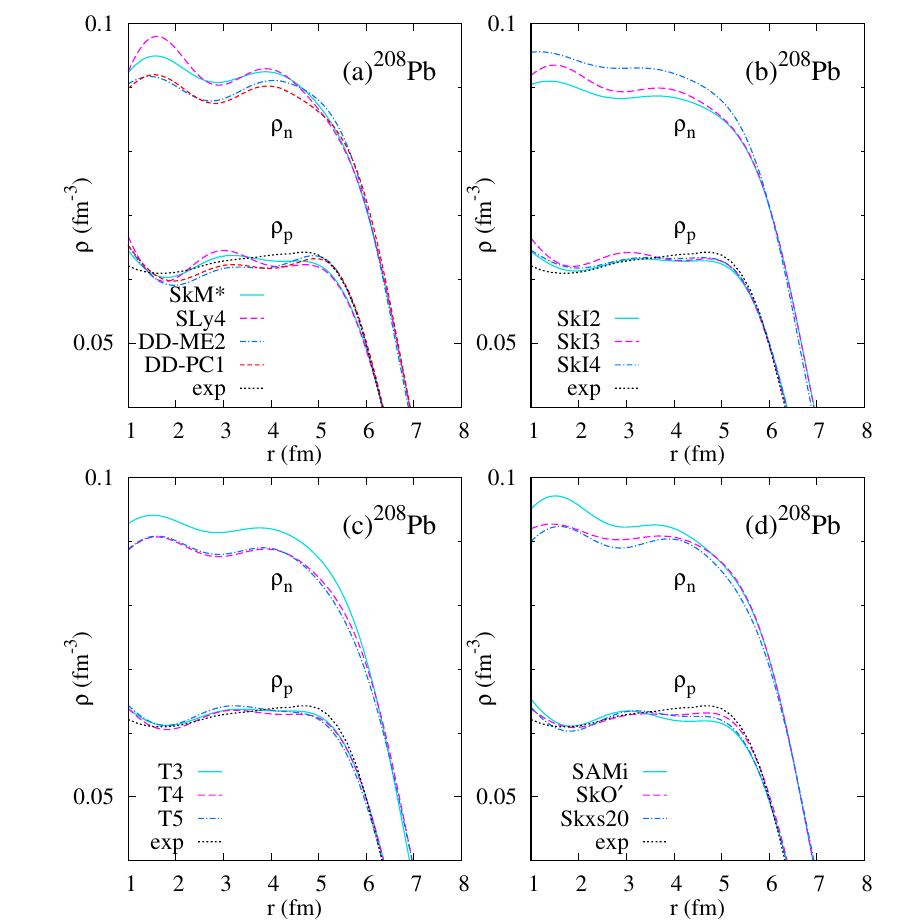}
\caption{Neutron [$\rho_n(r)$] and proton [$\rho_p(r)$] densities
of $\Pb208$ obtained from SHFB
calculations using (a) the SLy4 and SkM*; (b) the SkI2, SkI3, and SkI4; (c) the SkT2 (T2), SkT3 (T3), and SkT4 (T4); and (d) the SAMi, SkO$^\prime$, and Skxs20 interactions,
together with the experimental proton density determined from electron-scattering data \cite{DeJager:1987qc}.
(The experimental values are taken from Ref.~\cite{Zenihiro:2010zz}.)
The results obtained from RHB calculations using the DD--ME2 and DD--PC1 interactions are also presented in panel (a).
\label{fig:dens-pn-s}}
\end{figure}

\begin{figure*}[!ptb]
\includegraphics[width=16 cm]{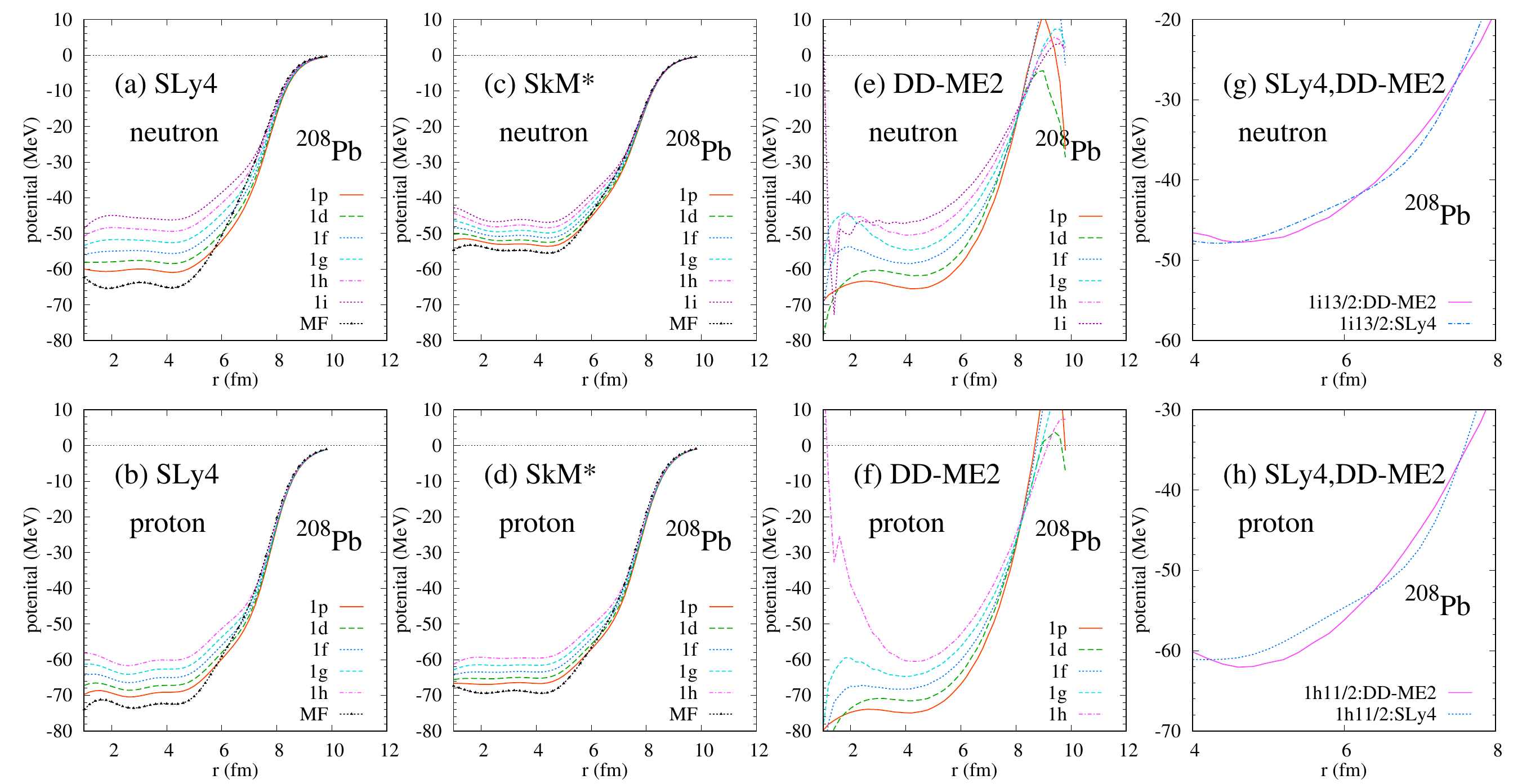}
\caption{
Averaged effective single-particle potentials $U^\textrm{eff}_{\alpha,\textrm{av}}(r)$ in $\Pb208$
obtained from SHFB calculations using the SLy4 interactions for (a) neutrons and (b) protons;
those obtained using SkM* for (c) neutrons and (d) protons; and the RHB results obtained using DD--ME2 for (e) neutrons and (f) protons.
The SLy4 and DD--ME2 results for neutrons are compared in panel (g) and those for protons are compared in
panel (h). The mean field $U^\textrm{MF}(r)$ is also displayed in panels (a), (b), (c), and (d).
\label{fig:pot-sp-av-pb208}}
\end{figure*}

\begin{figure}[!ptb]
\includegraphics[width=8.6 cm]{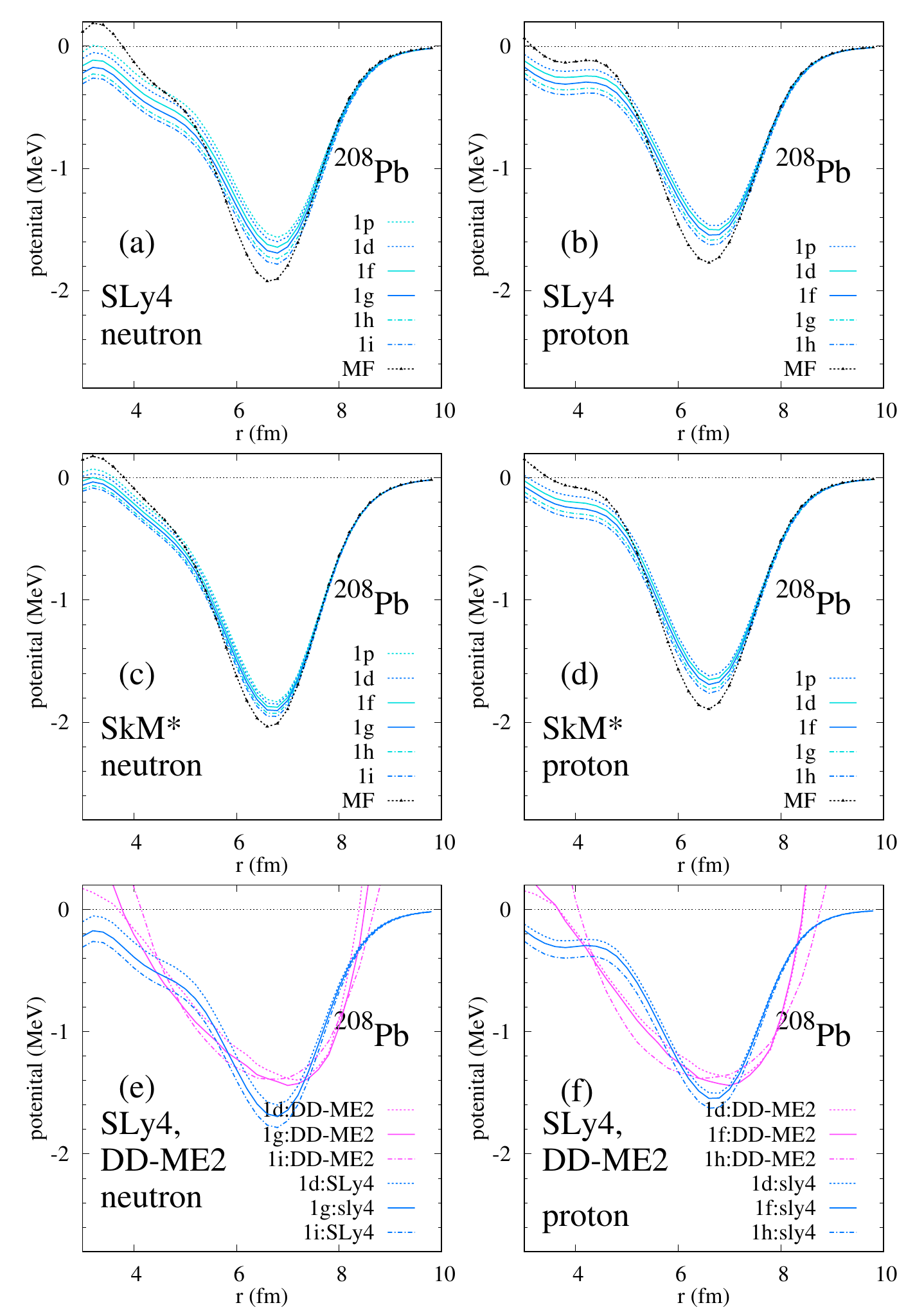}
\caption{The effective $\ell s$ potentials $U^{\textrm{eff}(\rho)}_{\ell s}$
in $\Pb208$ obtained from SHFB calculations with the SLy4 interaction for (a) neutrons and (b) protons;
those obtained with SkM* for (c) neutrons and (d) protons; and those obtained from RHB calculations
using the DD--ME2 interaction for (e) neutrons and (f) protons. In panels (e) and (f), the SLy4 results
are also presented for comparison.
The mean-field $\ell s$ potentials $U^\textrm{MF}_{\ell s}$ obtained in the SHFB results are also displayed in panels (a), (b), (c), and (d).
\label{fig:pot-sp-ls-me2}}
\end{figure}

\begin{figure}[!ptb]
\includegraphics[width=8.6 cm]{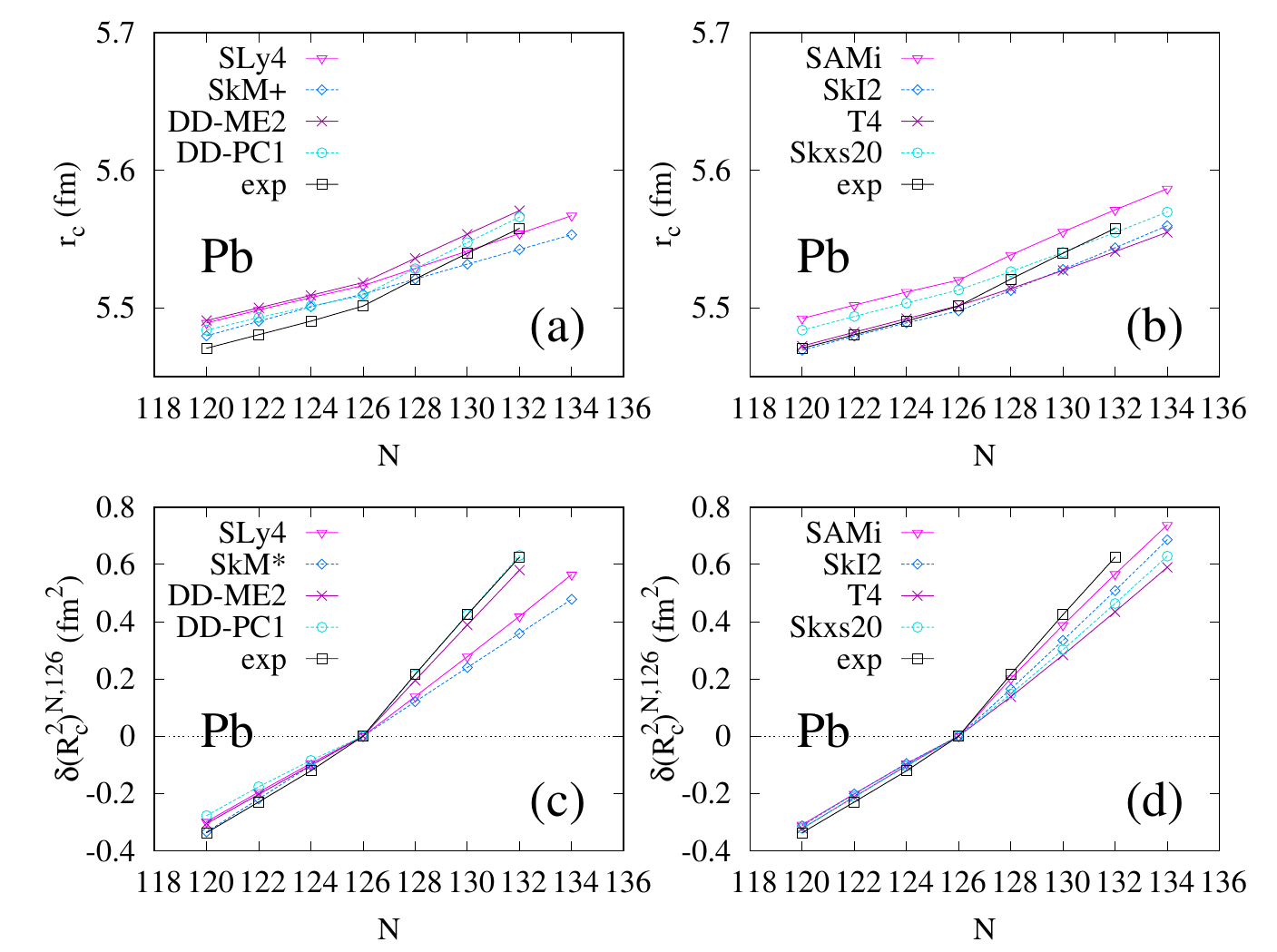}
\caption{ (a) Charge radii $r_c$ and (c) differential mean-square charge radii
$\delta(r^2_c)^{N,126}$ of Pb isotopes obtained
from SHFB and RHB calculations, and (b) $r_c$ and (d) $\delta(r^2_c)^{N,126}$
obtained from various SHFB calculations.
\label{fig:rc-pbiso}}
\end{figure}

\begin{figure}[!htpb]
\includegraphics[width=7.5 cm]{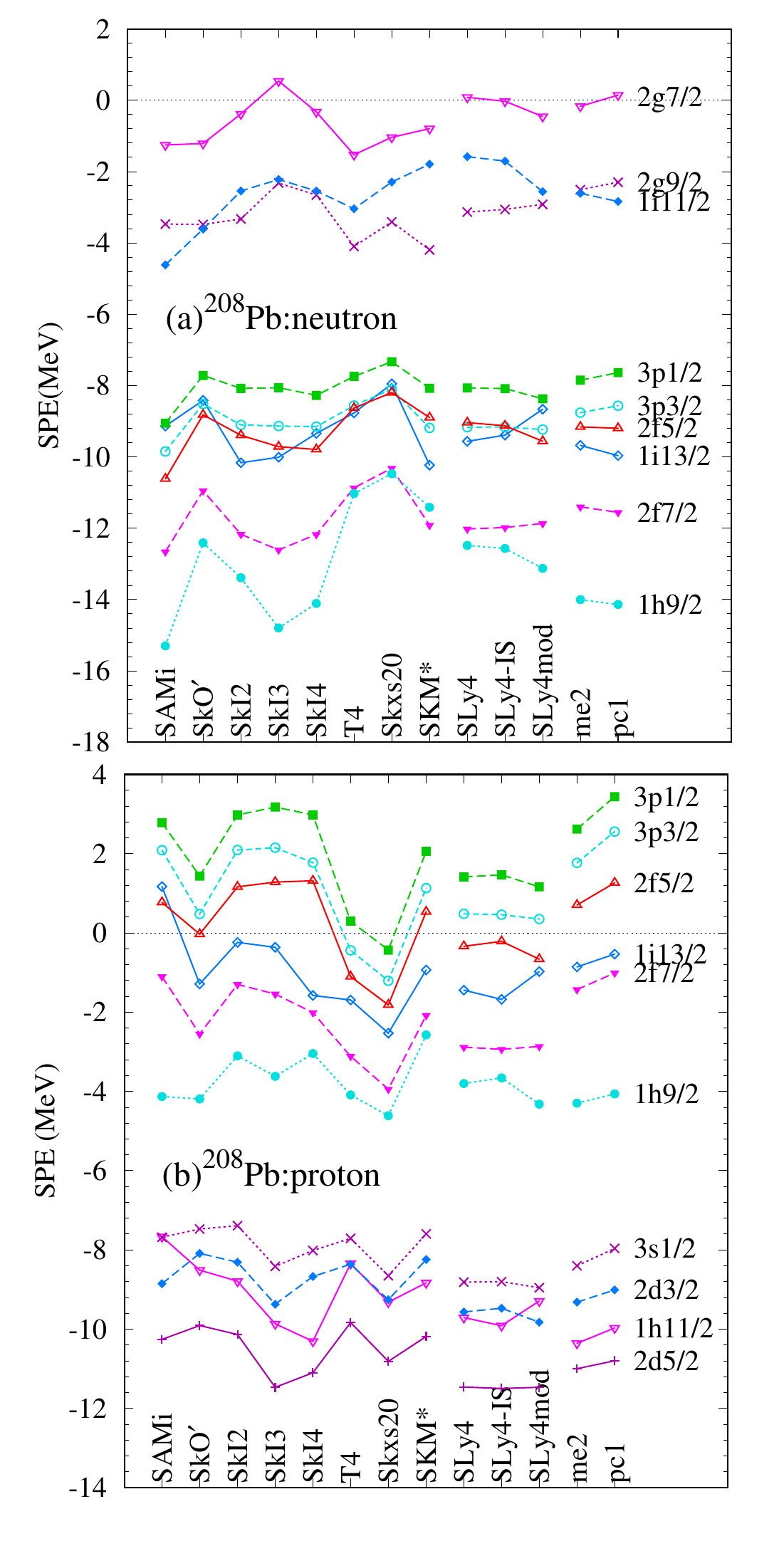}
\caption{Single-particle energies (SPEs) of (a) neutrons and (b) protons in $\Pb208$
obtained from SHFB and RHB calculations. The label "me2" represents the DD--ME2 interaction, and "pc1" similarly represents DD--PC1.
\label{fig:spe-pb208}}
\end{figure}

\begin{figure}[!htpb]
\includegraphics[width=8.6 cm]{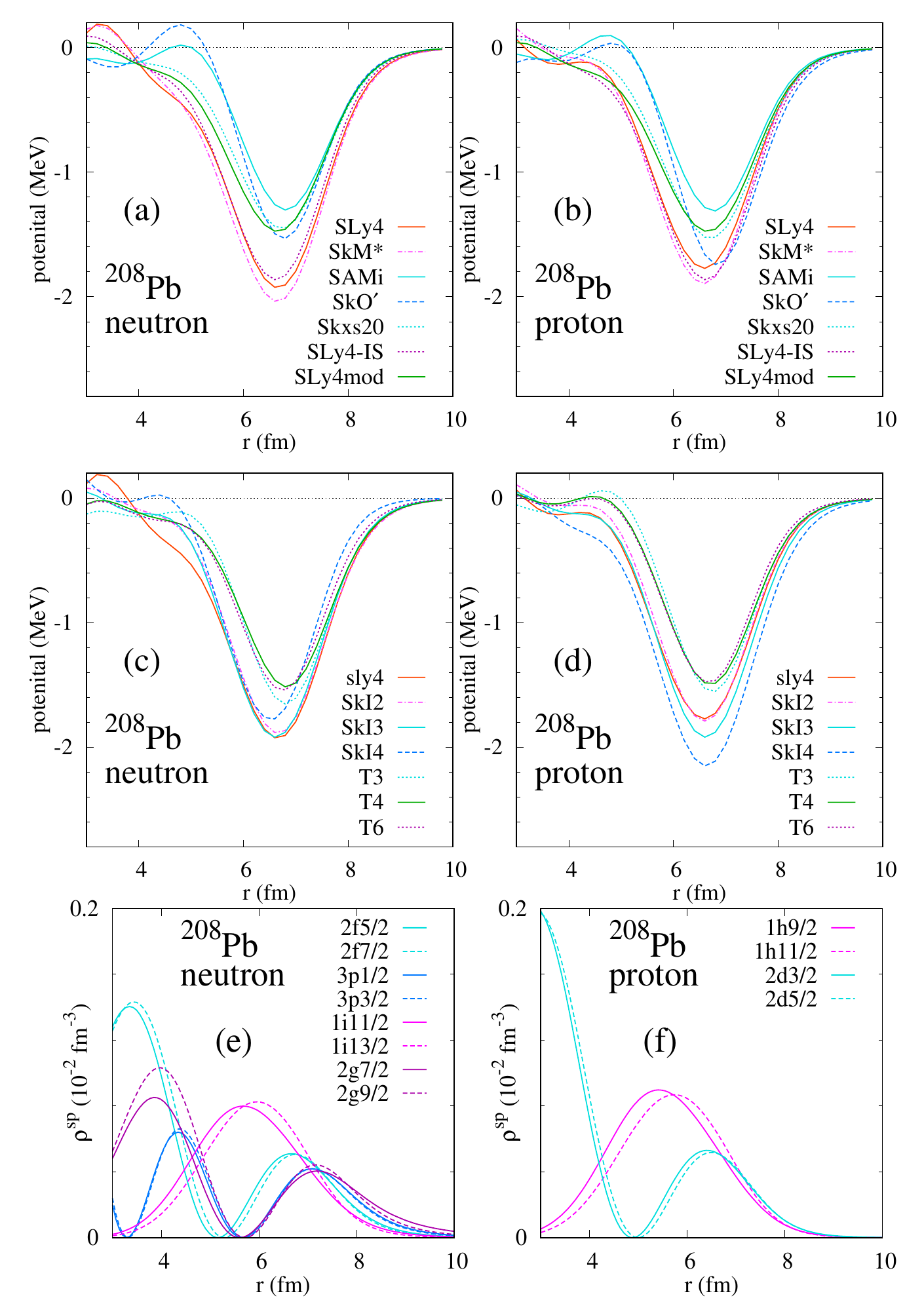}
\caption{(Top and middle panels) The mean-field $\ell s$ potentials $U^\textrm{MF}_{\ell s}$
and (bottom panels)
the single-particle densities of (left) neutrons and (right) protons in $\Pb208$
obtained using the Skryme interactions.
\label{fig:pot-pn-ls}}
\end{figure}

The neutron ($\rho_n)$ and proton ($\rho_p$) densities of $\Pb208$ obtained
from SHFB calculations
using the SLy4 and SkM* interactions and those obtained from
RHB calculations using the DD--ME2 and DD--PC1 interactions are presented
in Figs.~\ref{fig:dens-pn}(a) and (b), respectively.
The SLy4 and SkM* interactions yield similar density distributions.
Compared with the experimental densities,
the SLy4 and SkM* results underestimate $\rho_n$ and $\rho_p$
in the inner-surface region at $r\sim 5$ fm.
On the other hand, the RHB calculations reproduce
the inner-surface densities well, particularly in the results obtained
using the DD--ME2 interaction.

Figures~\ref{fig:dens-pn}(c) and (d), respectively, show the distributions
$\rho_n$ and $\rho_p$ obtained for$\Ca48$
from the SHFB and RHB calculations.
They show similar trends in the inner-surface densities;
the RHB results are in good agreement with the
experimental data, whereas the SLy4 and SkM* calculations yield smaller densities
than the RHB results in the inner-surface region at $r\sim 2.5$~fm.

Results obtained for the densities of $\Pb208$ using other Skyrme interactions
are presented in Fig.~\ref{fig:dens-pn-s}. The results for
$\rho_p$ at $r\sim 5$~fm in the inner-surface region obtained using
the SkI2, SAMi, and Skxs20 interactions are similar to those obtained using
SLy4 and SkM*.
The SkI3, SkI4, SkT3, SkT4, SkT5, and SkO$^\prime$ interactions yield
slightly better results, but they still underestimate the inner
surface proton density at $r\sim 5$~fm compared
with the experimental data.

To understand the different trends between the SHFB and RHB results for
the inner part of the surface proton density,
I calculated the effective single-particle potentials using the 
single-particle densities obtained
as explained in Sec.~\ref{subsec:eff-pot}.
The effective potentials $U^\textrm{eff}_{\alpha,\textrm{av}}(r)$
averaged over the $\alpha_>$ and $\alpha_<$ orbitals in $\Pb208$
are shown in Figs.~\ref{fig:pot-sp-av-pb208}(a)--(f). The effective potentials are $\ell$-dependent
due to the effective-mass contribution.
In addition to the SHFB results,
Figs.~\ref{fig:pot-sp-av-pb208}(a)--(d) display the mean-field (HF) potentials which do not contain the effective-mass contribution.
The effective mass $m^*$ in nuclear matter at normal density
is $m^*/m=0.572$ in the DD--ME2 case and $m^*/m=0.69$ in the SLy4 case ($0.79$ in the SkM* case).
The low-$\ell$ orbital potentials are deeper in the DD-ME result
because effective mass is smaller than in the
SLy4 and SkM* results, but the potential depths of high-$\ell$ orbitals are similar
for all three interactions (DD--ME2, SLy4, and SkM*).
Figure~\ref{fig:pot-sp-av-pb208}(g) [\ref{fig:pot-sp-av-pb208}(h)] compares the SLy4 and DD--ME2 results for $U^\textrm{eff}_{\alpha_>}((r)$ for the highest-$\ell$ orbital of the major-shell neutrons (protons).
In the inner-surface region at $r\sim 5$ fm,
the DD--ME2 interaction produces deeper potentials for the highest-$\ell$ orbitals.
In particular, the proton $1h_{11/2}$ potential is significantly deeper in DD--ME2.
Due to the deep potential, $\rho_p$ is increased at $r\sim 5$ in the DD--ME2 result
because it is dominated by contributions from the $1h_{11/2}$ orbital.
In other words, the reason why the SLy4 results underestimate the inner part of the surface proton density
is because
the effective potential at $r\sim 5$ is shallower than the 
DD--ME2 case.

The effective $\ell s$ potentials $U^\textrm{eff}_{\alpha,\ell s}$
in $\Pb208$ are presented in Fig.~\ref{fig:pot-sp-ls-me2}.
Figures \ref{fig:pot-sp-ls-me2}(e) and \ref{fig:pot-sp-ls-me2}(f) compare the
SLy4 and DD--ME2 results.
The $r$-dependences of the $\ell s$ potentials of these two results 
are qualitatively different.
In the inner part of the surface region at $r\sim 5$~fm,
the proton $\ell s$ potential is shallower in the SLy4 result
than in the DD--ME2 result.
On the other hand, in the outer part of the surface region at $r\sim 7$~fm,
the neutron $\ell s$ potential is significantly deeper in the
SLy4 result than the DD--ME2 case.

These differences in the surface potentials between the SLy4 and DD--ME2 results cause
quantitative differences in the surface densities, and they also
contribute to the SPEs, as discussed later in this paper.

\subsection{Charge radii of Pb isotopes}

To discuss the $N$ dependence of the charge radii of the Pb isotopes,
I define the differential mean-square charge radius as
\begin{align}
\delta(r_c^2)^{N,N'}=r_c^2(N)-r_c^2(N').
\end{align}
Here $r_c^2(N)$ represents the mean-square charge radius of a nucleus with neutron number $N$; 
$N'$ is the neutron number of the reference nucleus, which I chose to be $N'=126$ for the Pb isotopes; i.e., $\Pb208$.

The calculated values of $r_c$ and $\delta(r_c^2)^{N,126}$ obtained from the SHFB and RHB calculations are
presented in Fig.~\ref{fig:rc-pbiso}, where they are compared with the experimental data.
Fig.~\ref{fig:rc-pbiso}(c) shows that the DD--ME2 and DD--PC1 calculations reproduce the kink
at $N=126$ in the experimental values of
$\delta(r_c^2)^{N,126}$ well.
Conversely, the SLy4 and SkM* results fail to reproduce the kink behavior, although
the SHFB calculations with the SAMi and SkI2 interactions yield better results than do the
SLy4 and SkM* interactions [see Fig.~\ref{fig:rc-pbiso}(d)].

As discussed in many works using non-relativistic and relativistic mean-field calculations,
the kink behavior of the charge radii in the Pb isotopes
is correlated with the energy difference $e(1i_{11/2})-e(2g_{9/2})$ of the major-shell neutron orbitals;
the kink behavior of the charge radii is enhanced when $e(1i_{11/2})-e(2g_{9/2})$ is small
because the charge radius increases rapidly for $N>126$ due to
the increasing occupation of the neutron $1i_{11/2}$ orbital.

The SPEs of the neutron and proton orbitals in $\Pb208$ obtained from the SHFB and RHB calculations
are displayed in Fig.~\ref{fig:spe-pb208}.
In the SLy4 and SkM* cases, which show weak or no kinks in $\delta(r_c^2)^{N,126}$,
the energy of the neutron $1i_{11/2}$ orbital  is much higher than is the energy of the $2g_{9/2}$ orbital. On the other hand, in the SHFB calculations that employ the SAMi, SkO$^\prime$, SkI2, SkI3,
and SkI4 interactions and which exhibit
better results for the kink behavior,
the $1i_{11/2}$ orbital becomes degenerate with (or even lower than)
the $2g_{9/2}$ orbital.
In the DD--ME2 and DD--PC1 cases, which reproduce the $N$ dependence of
experimental results for $\delta(r_c^2)^{N,126}$ well, the $1i_{11/2}$ and $2g_{9/2}$ orbitals are almost
degenerate.

The $r$-dependence of the neutron $\ell s$ potential affects the energy difference $e(1i_{11/2})-e(2g_{9/2})$
through its contribution to
the $\ell s$ splittings $\Delta^\nu_{\ell s}(1i)\equiv e(1i_{11/2})-e(1i_{13/2})$
of the neutron $1i_{11/2,13/2}$ orbitals and $\Delta^\nu_{\ell s}(2g)\equiv e(2g_{7/2})-e(2g_{9/2})$
of the neutron $2g_{7/2,9/2}$ orbitals.
The mean-field $\ell s$ potentials ($U^\textrm{MF}_{\ell s}$) 
in $\Pb208$ obtained from the SHFB calculations
are presented in Fig.~\ref{fig:pot-pn-ls}, together with the
single-particle densities ($\rho^\textrm{sp}_\alpha$) of $\Pb208$ from the SLy4 results.
Here, the $\ell s$ potential $U^\textrm{MF}_{\ell s}$ is defined as the $r$-dependent part of the
spin--orbit potential of the mean field:
\begin{align}
U^\textrm{MF}=U_\textrm{cent}(r)+U^\textrm{MF}_{\ell s}\bvec{\ell}\cdot \bvec{s}.
\end{align}
The potential depth of $U^\textrm{MF}_{\ell s}$ in the region $r=5$--6~fm 
contributes to $\Delta^\nu_{\ell s}(1i)$
because the neutron $1i_{11/2,13/2}$ orbitals have peaks at $r=5$--6~fm.
On the other hand,
$U^\textrm{MF}_{\ell s}$ in the region $r\sim 7$~fm 
contributes to $\Delta^\nu_{\ell s}(2g)$
because the $2g_{7/2,9/2}$ orbitals have surface peak amplitudes
at $r\sim 7$~fm.
Compared with the SLy4 and SkM* interactions,
the SAMi, SkO$^\prime$, SkI2, and SkI3 interactions provide shallower ${\ell s}$ potentials for the neutrons
over the whole range of $r$. The shallower ${\ell s}$ potentials---in particular,
in the region $r=5$--6~fm---decrease the $\ell s$-splitting $\Delta^\nu_{\ell s}(1i)$
and lower the energy of the $1i_{11/2}$ orbital.
For the SkI4 interaction, the neutron $\ell s$ potential in the region $r=5$--6~fm 
is comparable to that obtained from the SLy4 and SkM* results,
but it is weaker than that in the region $r\sim 7$~fm; 
it therefore decreases $\Delta^\nu_{\ell s}(2g)$ and raises the $2g_{9/2}$ energy.

Let us next consider the $\ell s$ splittings in the DD--ME2 and DD--PC1 results.
As shown in Fig.~\ref{fig:spe-pb208}(a), the DD--ME2 and DD--PC1 results
exhibit smaller $\ell s$ splittings of the neutron $1i_{11/2,13/2}$ orbitals
and $2g_{7/2,9/2}$ orbitals than do the SLy4 and SkM* results.
In particular, the $\ell s$-splitting of the $2g_{7/2,9/2}$ orbitals
is significantly smaller than that obtained with the SLy4 and SkM* interactions
because the $\ell s$ potentials are shallower in the region $r\sim 7$~fm, 
as shown in Fig.~\ref{fig:pot-sp-ls-me2}(e).

Refs.~\cite{Reinhard:1995zz,Goddard:2012dk} argued that
the IS/IV ratio of the spin--orbit potentials plays an important role in the kink phenomenon.
The SLy4, SkM*, SkI2, and Skxs20 interactions use the usual ratio ($b_4=b^\prime_4$), 
while the SkI3 and SAMi interactions 
employ IS-type ($b^\prime_4=0$) and IS-dominant spin--orbit terms, respectively.
The SkI4 and SkO$^\prime$ interactions employ $b_4\sim -b_4'$
 for the reverse-type ratio, but there is no fundamental reason. Assuming $b_4\sim -b_4'$
means that the neutron $\ell s$ potential is determined only
by the proton density---as it is $\propto\frac{1}{r}\frac{d\rho_p}{dr}$---
which seems unrealistic.

To investigate the contribution of the IS/IV ratio in the spin--orbit term
to the SPEs and $\ell s$ potentials in the SHFB calculations,
I compared the results obtained using the SLy4 and SLy4--IS interactions,
which employ usual-type and IS-type spin--orbit terms, respectively,
but leaving the other parameters unchanged.
The calculated SPEs and $\ell s$ potentials are shown 
in Figs.~\ref{fig:spe-pb208}(a) and \ref{fig:pot-pn-ls}(a) for neutrons and
in Figs.~\ref{fig:spe-pb208}(b) and \ref{fig:pot-pn-ls}(b) for protons.
There are minor differences in the SPEs between the SLy4 and SLy4--IS results, although the $\ell s$ potential for the neutrons (protons) is
slightly weaker (stronger) in the SLy4--IS result than in the SLy4 result.
This indicates that the SPEs are not significantly
affected by the IS/IV ratio of the spin--orbit interactions.

In Figs.~\ref{fig:spe-pb208} and \ref{fig:pot-pn-ls},
I also present the results obtained using the SLy4mod interaction.
This interaction is another version, with the IS-type spin--orbit term
modified from the SLy4, and
all the parameters are consistent with those of the SLy4--IS interactions
except for the value of $b_4$.
The SLy4mod interaction employs $b_4=75$~MeV and $b^\prime_4=0$ corresponding to a 
spin--orbit interaction that is
20\% weaker 
than are those in the SLy4 and SLy4--IS interactions.
Due to the weaker spin--orbit term,
the neutron $1i_{11/2}$ and $2g_{9/2}$ orbitals are almost degenerate
with each other [Figs.~\ref{fig:spe-pb208} and \ref{fig:pot-pn-ls}(a)],
and the SLy4mod result consequently gives better results for the kink in the charge radii of the Pb isotopes.
It should be stressed that a weaker spin--orbit term
is essential for reproducing the kink behavior
in the SLy4mod result, but the role of the IS/IV ratio
is minor.

\section{Results from the new Skyrme interactions with a density-dependent spin--orbit term}\label{sec:new-results}

As discussed previously, the IS/IV ratio of the spin--orbit interactions
produces only minor effects in the SHFB results for the SPEs and charge radii of the Pb isotopes.
Instead, other extensions beyond the conventional Skyrme EDF need to be considered
to improve the results for this kink behavior.
Comparing the DD--ME2 and SLy4 results shows that the SLy4 calculation has the problems 
of overestimating the energy difference $e(1h_{11/2})-e(2g_{9/2})$ of the neutrons and
underestimating the inner part of the surface proton density. These problems may arise from 
defect in the $r$-dependence of the $\ell s$ potentials in $\Pb208$;
compared with the DD--ME2 case,
the SLy4 interaction yields
a deeper neutron $\ell s$ potential in the region $r\sim 7$~fm [Fig.~\ref{fig:pot-sp-ls-me2}(e)],
which increases $e(1h_{11/2})-e(2g_{9/2})$ for the neutrons by reducing the $\ell s $ splitting $\Delta^\nu_{\ell s}(2g)$.
Furthermore, the SLy4 interaction produces a shallower proton $\ell s$ potential in the region $r=5$--6~fm 
[Fig.~\ref{fig:pot-sp-ls-me2}(f)],
which decreases the proton density at $r\sim 5$~fm.

To modify the $r$-dependence of the $\ell s$ potentials in the SHFB framework, I introduced a 
density-dependent strength for the spin--orbit term and constructed new interactions, called
Skyrme--ddso, as explained in Sec.~\ref{sec:calculations}.
I then compared the results obtained from the SHFB calculations
using the Skyrme--ddso interactions with the original Skyrme results
to see how the kink phenomenon of the charge radii and the surface proton densities are affected by
the introduction of the density-dependent spin--orbit term.

\subsection{Binding energies and radii}

The SLy4 and SLy4--ddso results for the binding energies per nucleon of doubly magic and
proton magic nuclei are listed in Table \ref{tab:be-sly4}, and those for the neutron, proton, and charge radii are listed in Table \ref{tab:rmsr}.
The SLy4--ddso calculation obtains
slightly smaller values for the binding energies and charge radii, but the differences
from the original SLy4 values are less than 0.6\% in the binding energies
and less than 0.8\% in the charge radii.
The SLy4--ddso2 results are quite similar to the
SLy4--ddso results, as shown in Table \ref{tab:be-sly4}.

\begin{figure}[!htpb]
\includegraphics[width=8.6 cm]{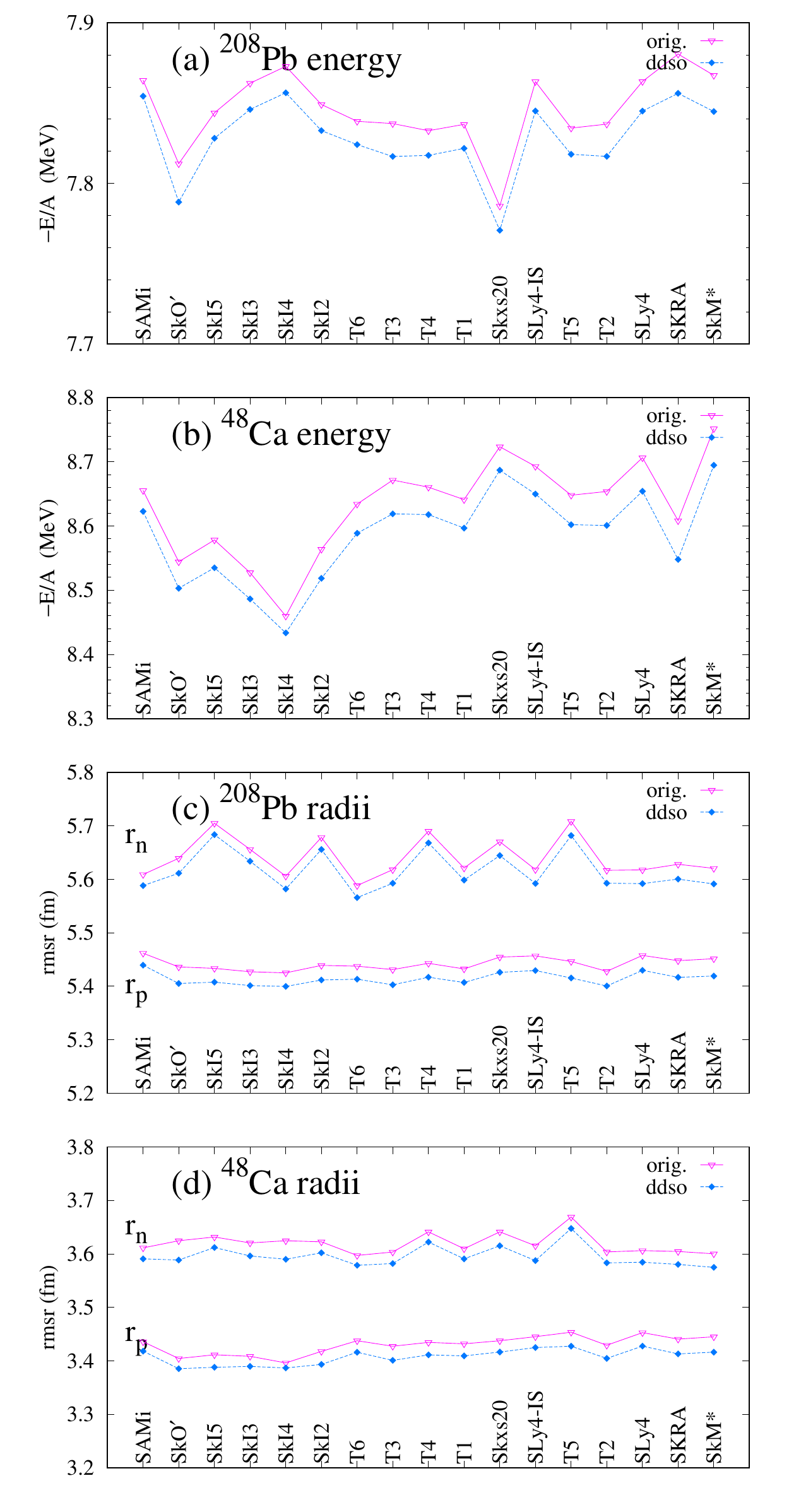}
\caption{
Binding energies per nucleon ($-E/A$) and rms neutron ($r_n$) and proton ($r_p$) radii
of $\Pb208$ and $\Ca48$ obtained from SHF calculations
using the Skyrme~(orig.) and Skyrme--ddso~(ddso) interactions.
\label{fig:rmsr-ene}}
\end{figure}

For other series of Skyrme interactions,
the Skyrme and Skyrme--ddso results for $\Pb208$ and $\Ca48$
are presented in Fig.~\ref{fig:rmsr-ene}.
For those Skyrme interactions also,
the inclusion of a density-independent spin--orbit term
yields little changes from the original Skyrme results.

\begin{table}[!htpb]
\caption{
Binding energies per nucleon ($-E/A$) obtained from SHFB calculations
using the SLy4, SLy4--ddso, and SLy4--ddso2 interactions.
The neutron-pairing energies of $\Ni60$, $\Sn120$, and $\Pb214$ are also listed.
\label{tab:be-sly4}
}
\begin{center}
\begin{tabular}{ccccccccccccccc}
\hline
\hline
&	\multicolumn{4}{c}{$-E/A$~(MeV)}& \multicolumn{3}{c}{pairing energy~(MeV)} \\ &	SLy4	&	SLy4	-	&	SLy4- &	exp	&	SLy4	&	SLy4	-	&	SLy4- \\
&	(orig)	&	ddso	&	ddso2	& &	(orig)	&	ddso	&	ddso2 \\
$\O16$	&	8.031 &	8.051 &	8.051 &	7.976 & & & \\
$\Ca40$	&	8.606 &	8.621 &	8.620 &	8.551 & & & \\
$\Ca48$	&	8.706 &	8.654 &	8.662 &	8.667 & & & \\
$\Ca52$	&	8.453 &	8.422 &	8.435 &	8.429 & & & \\
$\Ni56$	&	8.631 &	8.605 &	8.598 &	8.643 & & & \\
$\Ni60$	&	8.776 &	8.766 &	8.761 &	8.781 & & & \\
$\Ni68$	&	8.713 &	8.718 &	8.706 &	8.682 & & & \\
$\Sr88$	&	8.736 &	8.698 &	8.703 &	8.733 & & & \\
$\Zr90$	&	8.730 &	8.700 &	8.705 &	8.710 & & & \\
$\Pb208$	&	7.864 &	7.845 &	7.845 &	7.867 & & & \\
$\Ni58$	&	8.719 &	8.697 &	8.691 &	8.732 &$	-2.609 $&$	-2.508 $&$	-2.526 $\\
$\Sn120$	&	8.490 &	8.480 &	8.473 &	8.504 &$	-12.570 $&$	-9.029 $&$	-8.069 $\\
$\Pb214$	&	7.744 &	7.720 &	7.721 &	7.772 &$	-11.467 $&$	-12.136 $&$	-12.773 $\\
\hline
\hline
\end{tabular}
\end{center}
\end{table}

\begin{table}[!htpb]
\caption{
The root-mean-square neutron ($r_n$), proton ($r_p$), and charge ($r_c$) radii obtained from SHFB calculations
using the SLy4, SLy4--ddso, and SLy4--ddso2 interactions,
together with the experimental values of $r_c$~\cite{Angeli:2013epw}.
\label{tab:rmsr}
}
\begin{center}
\begin{tabular}{cccccccccccccc}
\hline
\hline
& \multicolumn{3}{c}{SLy4} & \multicolumn{3}{c}{SLy4-ddso} & exp \\
&	$r_n$	&	$r_p$	&	$r_c$	&	$r_n$	&	$r_p$	&	$r_c$	&	$r_c$	\\
$\O16$	&	2.661 &	2.686 &	2.803 &	2.655 &	2.680 &	2.797 &	2.699	\\
$\Ca40$	&	3.372 &	3.420 &	3.512 &	3.367 &	3.415 &	3.507 &	3.478	\\
$\Ca48$	&	3.606 &	3.453 &	3.544 &	3.585 &	3.428 &	3.520 &	3.477	\\
$\Ca52$	&	3.777 &	3.493 &	3.584 &	3.746 &	3.463 &	3.554 &	3.553	\\
$\Ni56$	&	3.647 &	3.702 &	3.787 &	3.595 &	3.649 &	3.736 & \\
$\Ni60$	&	3.770 &	3.730 &	3.815 &	3.718 &	3.676 &	3.762 &	3.812	\\
$\Ni68$	&	4.012 &	3.839 &	3.921 &	3.990 &	3.814 &	3.897 & \\
$\Sr88$	&	4.274 &	4.179 &	4.255 &	4.254 &	4.158 &	4.234 &	4.224	\\
$\Zr90$	&	4.288 &	4.225 &	4.300 &	4.269 &	4.206 &	4.281 &	4.269	\\
$\Pb208$	&	5.617 &	5.458 &	5.516 &	5.592 &	5.430 &	5.488 &	5.501	\\
$\Ni58$	&	4.730 &	4.594 &	4.663 &	4.715 &	4.574 &	4.643 &	3.776	\\
$\Sn120$	&	3.712 &	3.717 &	3.803 &	3.663 &	3.666 &	3.752 &	4.652	\\
$\Pb214$	&	5.695 &	5.496 &	5.554 &	5.673 &	5.471 &	5.529 &	5.557	\\
\hline
\hline
\end{tabular}
\end{center}
\end{table}

\subsection{Effects on the SPE of $\Pb208$}

\begin{figure}[!htpb]
\includegraphics[width=8.6 cm]{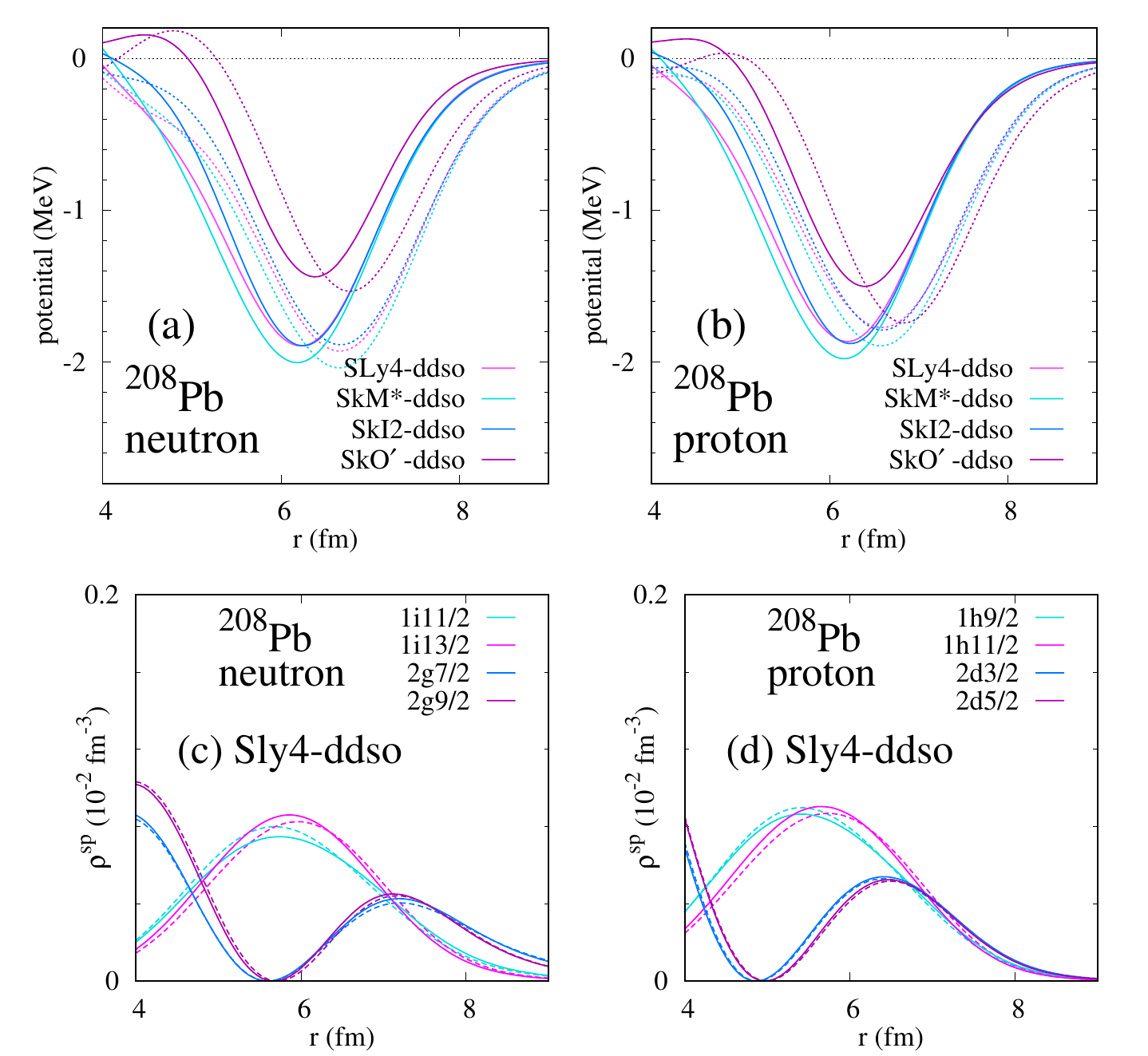}
\caption{
The Skyrme--ddso results (solid lines) for the mean-field $\ell s$ potentials ($U^\textrm{MF}_{\ell s}$) in $\Pb208$
for (a) neutrons and (b) protons. The lower panels show the SLy4--ddso
results (solid lines) for the single-particle densities
for (c) neutrons and (d) protons compared with the original
Skyrme results (dotted lines).
\label{fig:pot-ddso}}
\end{figure}

\begin{figure}[!htpb]
\includegraphics[width=8.6 cm]{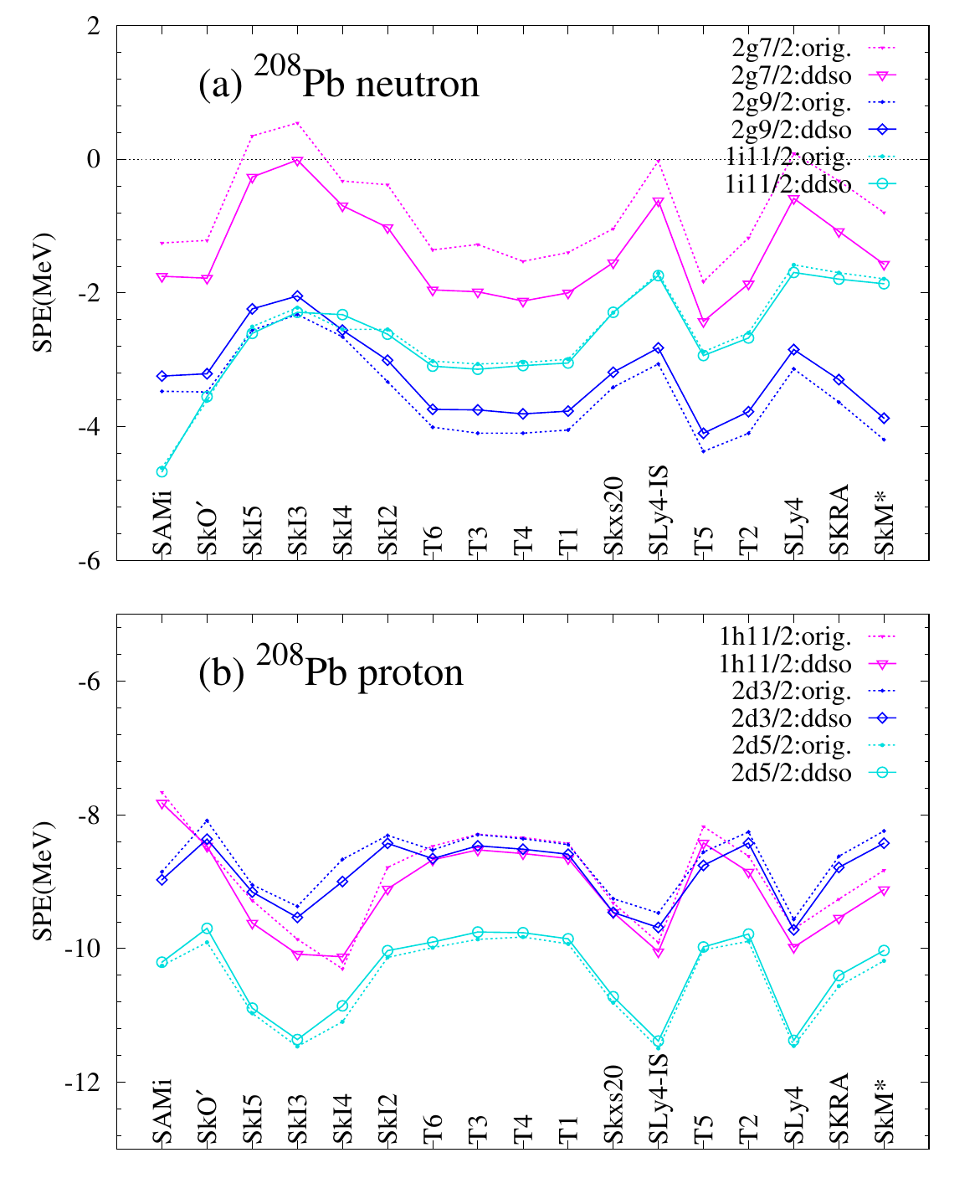}
\caption{(a) Neutron and (b) proton single-particle energies in
$\Pb208$ calculated using the Skyrme~(orig.) and Skyrme--ddso~(ddso) interactions.
\label{fig:spe-ddso-pb208}}
\end{figure}

\begin{figure}[!htpb]
\includegraphics[width=8.6 cm]{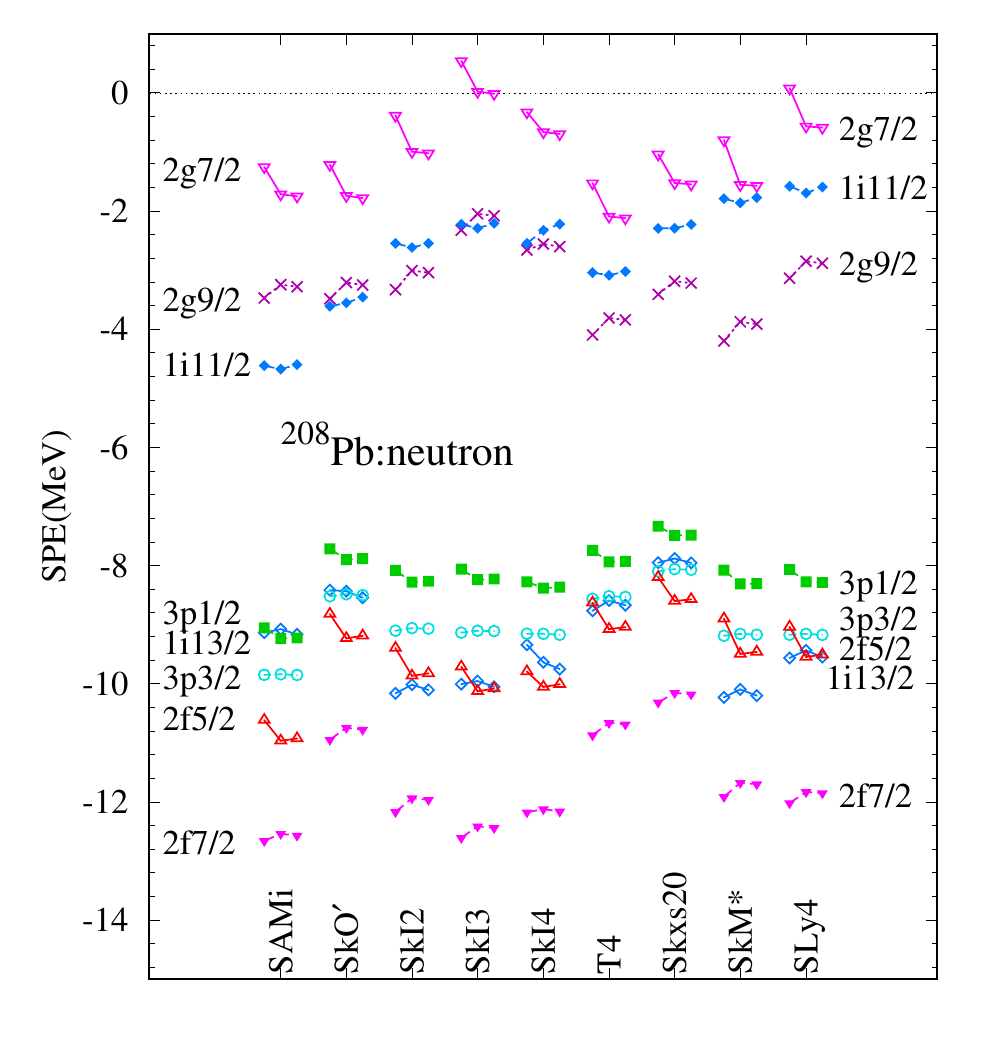}
\caption{Single-particle energies of neutrons in
$\Pb208$ calculated using the Skyrme, Skyrme--ddso, and Skyrme--ddso2 interactions.
In each set of three connected points, the values of
the Skyrme, Skyrme--ddso, and Skyrme--ddso2 results are plotted
on the left, middle, and right, respectively.
\label{fig:spe-ddso-pb208-2}}
\end{figure}

Figure~\ref{fig:pot-ddso} compares
the Skyrme and Skyrme--ddso results for
the $\ell s$ potentials ($U^\textrm{MF}_{\ell s}$) and the single-particle densities ($\rho^\textrm{sp}_\alpha$)
of neutrons and protons
in $\Pb208$, and Fig.~\ref{fig:spe-ddso-pb208} presents the corresponding results for the SPEs.
In all the Skyrme interactions,
the potential pocket of $U^\textrm{MF}_{\ell s}$ is shifted inward by the inclusion of the density-dependent spin--orbit term in the Skyrme--ddso results [Figs.~\ref{fig:pot-ddso}(a) and (b)].
The neutron $\ell s$ potential is comparable with the original Skyrme result in the region
$r\sim 6.5$~fm because I adjusted the spin--orbit strength ($w_4$) in the Skyrme-ddso interaction
to obtain the original value of the spin--orbit energy, but it is
significantly reduced in the outer region near $r\sim 7$~fm.
This reduction of the neutron $\ell s$ potential in the region $r\sim 7$~fm 
affects the neutron SPEs;
it decreases the $\ell s$ splitting $\Delta^\nu_{\ell s}(2g)$ because the neutron $2g_{7/2,9/2}$ orbitals
have surface amplitudes in this region, but
the $\ell s$ splitting $\Delta^\nu_{\ell s}(1i)$ is not affected
because the neutron $1h_{11/2,13/2}$ orbitals have no peak in this region but 
have significant amplitudes in the region $r\sim 6$~fm.
As a result of the reduction of $\Delta^\nu_{\ell s}(2g)$,
the energy difference $e(1i_{11/2})-e(2g_{9/2})$ is decreased by the inclusion of the
density-dependent spin--orbit term in the SLy4--ddso results [Fig.~\ref{fig:spe-ddso-pb208}(a)].
The effects on the proton SPEs of including the density-dependent spin--orbit term
are not as significant as they are for the neutrons
[Fig.~\ref{fig:spe-ddso-pb208}(b)], although this term does 
affect the
proton $\ell s$ potentials somewhat [Fig.~\ref{fig:pot-ddso}(b)].

To see the contribution of the IS/IV ratio of the density-dependent
spin--orbit part, in Fig.~\ref{fig:spe-ddso-pb208-2}
 I compare
the Skyrme--ddso2 results for
the neutron SPEs in $\Pb208$ 
with the Skyrme--ddso results.
There is no qualitative difference in the neutron SPEs between the Skyrme-ddso results with 
 $x_4=1$ (IS-type) and the Skyrme--ddso2 results with $x_4=0.5$ (usual-type), 
meaning that the choice of the IS/IV ratio does not significantly affect the SPEs.

\subsection{Kink phenomena in the Pb isotopes}

\begin{figure}[!htpb]
\includegraphics[width=8.6 cm]{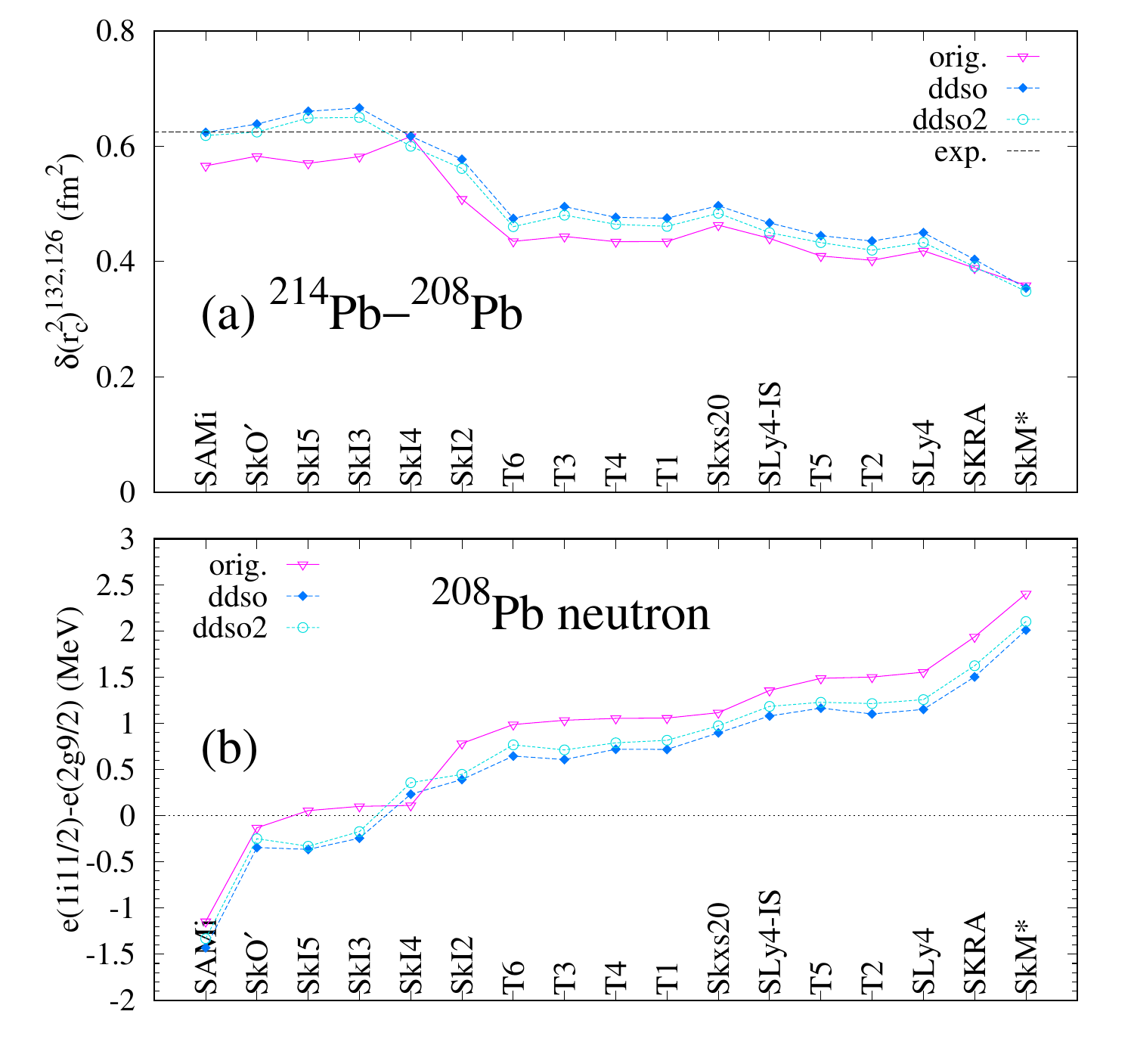}
\caption{(a) Differential mean-square charge radius $\delta(r^2_c)^{132,126}$
of the $\Pb214$--$\Pb208$ difference,
and (b) the energy differences $e(1i_{11/2})-e(2g_{9/2})$ of neutrons
in $\Pb208$
calculated using the Skyrme (orig.), Skyrme--ddso (ddso), and Skyrme--ddso2 (ddso2) interactions.
\label{fig:r2-pb208}}
\end{figure}

\begin{figure}[!htpb]
\includegraphics[width=8.6 cm]{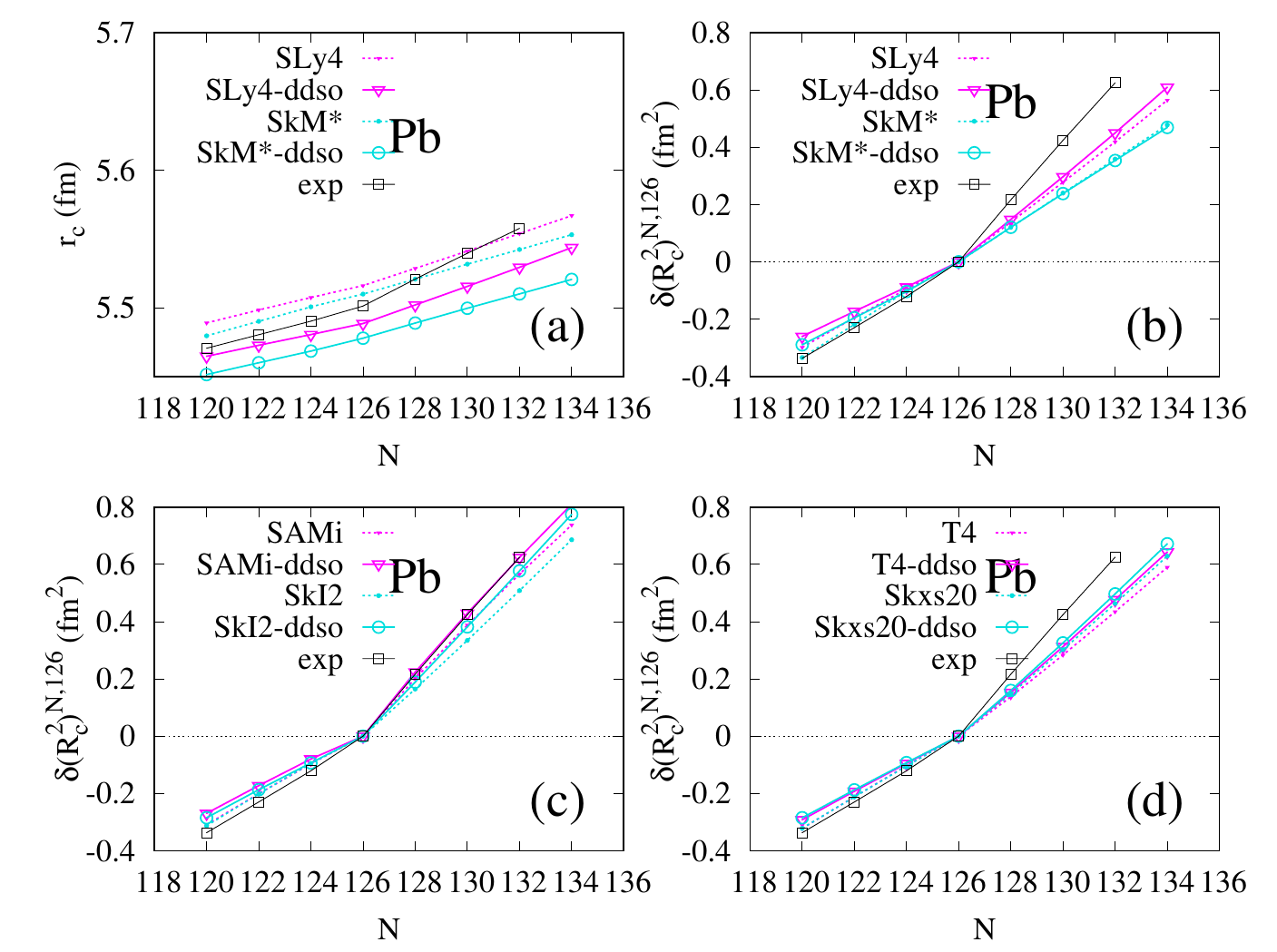}
\caption{(a) Charge radii $r_c$ and (b,c,d) differential mean-square charge radii $\delta(r^2_c)^{N,126}$
For the Pb isotopes obtained from SHFB calculations using the Skyrme and Skyrme--ddso interactions.
\label{fig:rc-ddso-pbiso}}
\end{figure}

The inclusion of the density-dependent spin--orbit term in the Skyrme--ddso interactions
affects the kink in $\delta(r_c^2)^{N,126}$ in the Pb isotopes through the decrease in the energy difference
$e(1i_{11/2})- e(2g_{9/2})$.
To see these effects, the calculated values of $\delta(r_c^2)^{132,126}$ for $\Pb214$ and of
$e(1i_{11/2})- e(2g_{9/2})$ for $\Pb208$
are presented in Figs.~\ref{fig:r2-pb208}(a) and \ref{fig:r2-pb208}(b), respectively.
In these figures,
the results of various Skyrme interactions are sorted by the $e(1i_{11/2})- e(2g_{9/2})$ values.
In the original Skyrme results, one can see
a clear correlation between $\delta(r_c^2)^{132,126}$ and
$e(1i_{11/2})- e(2g_{9/2})$; in general, larger values of $\delta(r_c^2)^{132,126}$ are obtained 
in cases with smaller values of $e(1i_{11/2})- e(2g_{9/2})$.
In the Skyrme--ddso results, $e(1i_{11/2})- e(2g_{9/2})$ is decreased by the inclusion of the density-dependent spin--orbit term,
yielding a larger value for $\delta(r_c^2)^{132,126}$ than in the original Skyrme results.

Figure~\ref{fig:rc-ddso-pbiso} compares the Skyrme and Skyrme--ddso results for $r_c$ and $\delta(r_c^2)^{N,126}$ for the
Pb isotopes. The Skyrme--ddso interactions yield better results for the kink behavior of $\delta(r_c^2)^{N,126}$
than do the original Skyrme results.
For instance, the SAMi--ddso and SkI2--ddso results are in good agreement with the experimental data for
$\delta(r_c^2)^{N,126}$. On the other hand, for the SLy4--ddso and Skxs20--ddso interactions,
 The improvements are not large enough to reproduce the kink behavior.

To show the contribution of the IS/IV ratio of the density-dependent
spin--orbit term,
the Skyrme--ddso2 results for $\delta(r_c^2)^{132,126}$ in the Pb isotopes and for $e(1i_{11/2})- e(2g_{9/2})$ in $\Pb208$ are compared with the Skyrme--ddso results in Fig.~\ref{fig:r2-pb208}.
The two calculations use the Skyrme--ddso interaction with $x_4=1$ (IS-type) and the Skyrme--ddso2 interaction with $x_4=0.5$ (usual-type), and they yield
qualitatively similar results for
$\delta(r_c^2)^{132,126}$ and for $e(1i_{11/2})- e(2g_{9/2})$,
although the Skyrme--ddso results yield quantitatively better results 
for $\delta(r_c^2)^{132,126}$.
This indicates that the choice of the IS/IV ratio 
for the density-dependent spin--orbit term causes only a minor difference in the kink phenomenon
in the Pb isotopes.

\subsection{Effects on the surface densities of $\Pb208$}

\begin{figure}[!htpb]
\includegraphics[width=8.6 cm]{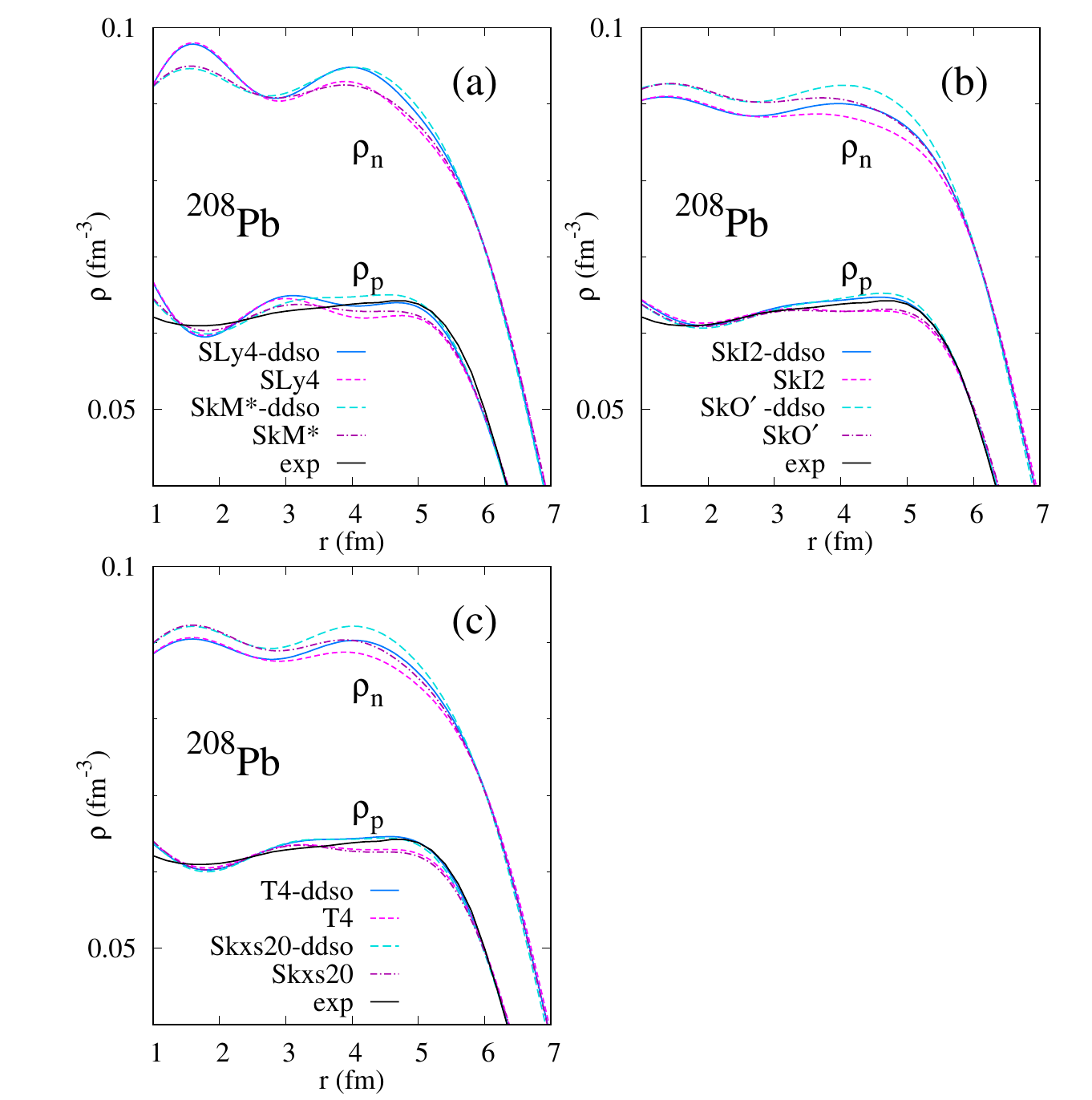}
\caption{Neutron and proton densities of $\Pb208$ obtained using the Skyrme and Skyrme--ddso interactions,
together with the experimental proton density determined from electron-scattering data \cite{DeJager:1987qc}.
(The experimental values are taken from Ref.~\cite{Zenihiro:2010zz}.)
\label{fig:dens-ddso-pn}}
\end{figure}

The neutron and proton densities of $\Pb208$ obtained using the Skyrme--ddso interactions
are compared with the original Skyrme results in Fig.~\ref{fig:dens-ddso-pn}.
The surface densities of $\Pb208$ are affected by the
the density-dependent spin--orbit term
through the change in the $r$-dependence of the $\ell s$ potentials.
In the Skyrme--ddso results, the inner-surface proton density at $r\sim 5$~fm is increased
mainly because the peak amplitude of the proton $1h_{11/2}$ orbital [Fig.~\ref{fig:pot-ddso}(d)]
is increased due to the deeper $\ell s$ potential in this region [Fig.~\ref{fig:pot-ddso}(b)].
As a result, the underestimate of the
surface proton density in the original Skyrme results is
improved by the inclusion of the density-dependent spin--orbit term, which produces good agreement between the experimental data and the Skyrme--ddso results.
The neutron density at $r\sim 5$~fm is also increased
because the peak amplitude of the neutron $1i_{13/2}$ orbital is increased
due to the deeper $\ell s$ potential in this region [Figs.~\ref{fig:pot-ddso}(a) and \ref{fig:pot-ddso}(c)].

\subsection{Results for $\Ca48$}

Next, I compared the results obtained from the SHF calculations for $\Ca48$
using the Skyrme--ddso interactions with the original Skyrme results
to discuss the effects of the density-dependent spin--orbit term in $\Ca48$.

\subsubsection{The SPEs and densities of $\Ca48$}

\begin{figure}[!htpb]
\includegraphics[width=8.6 cm]{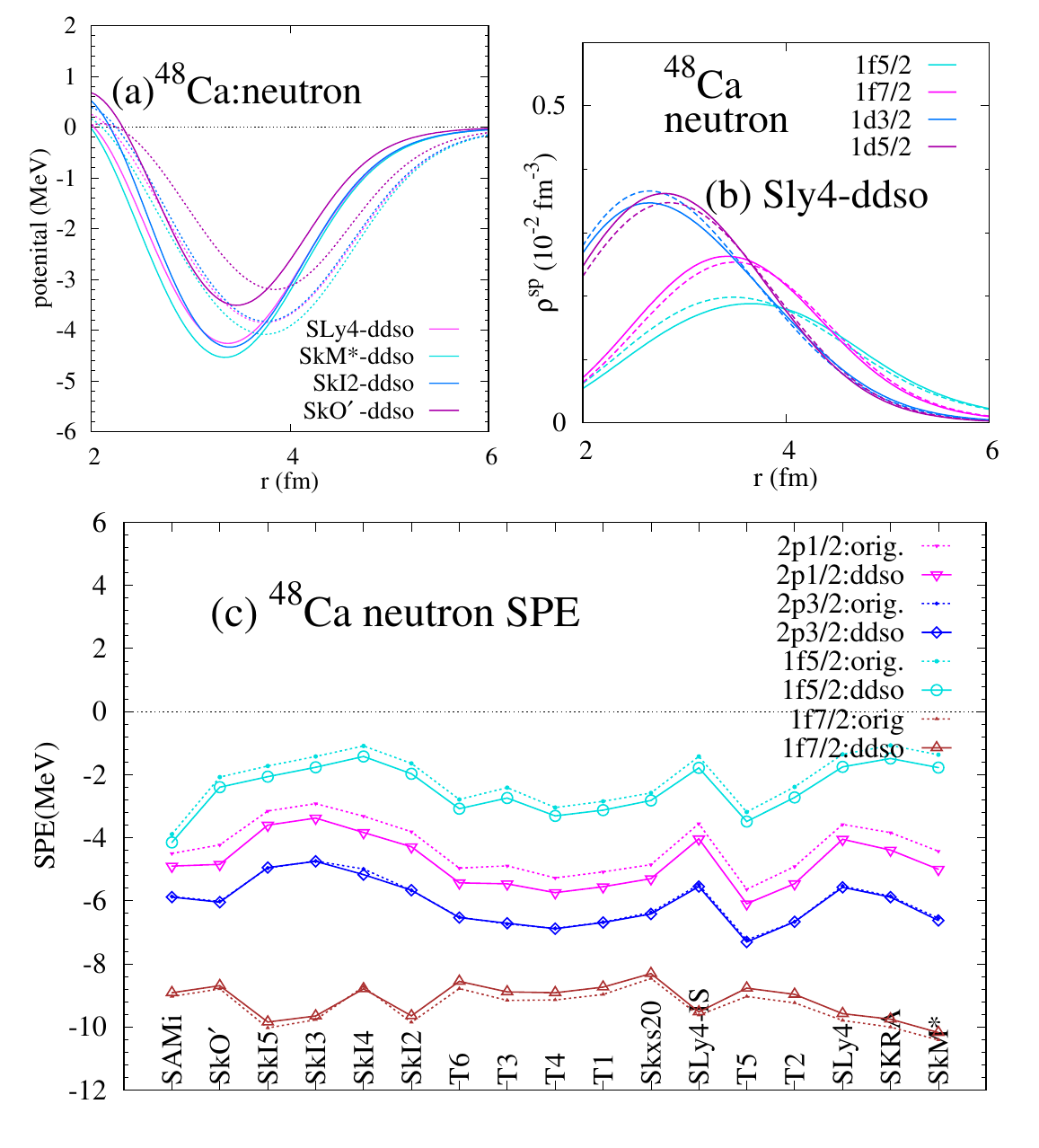}
\caption{
(a) The mean-field $\ell s$ potentials $U^\textrm{MF}_{\ell s}$ 
and (b) the single-particle neutron densities in $\Ca48$
obtained using the SLy4--ddso and SkM*--ddso interactions (solid lines)
compared with the original Skyrme interactions (dotted lines).
(c) The single-particle neutron energies (SPEs) in $\Ca48$ obtained from SHF calculations
using the Skyrme (orig.:dotted lines) and Skyrme--ddso (ddso:solid lines) interactions.
\label{fig:pot-ddso-ca48}}
\end{figure}

\begin{figure}[!htpb]
\includegraphics[width=8.6 cm]{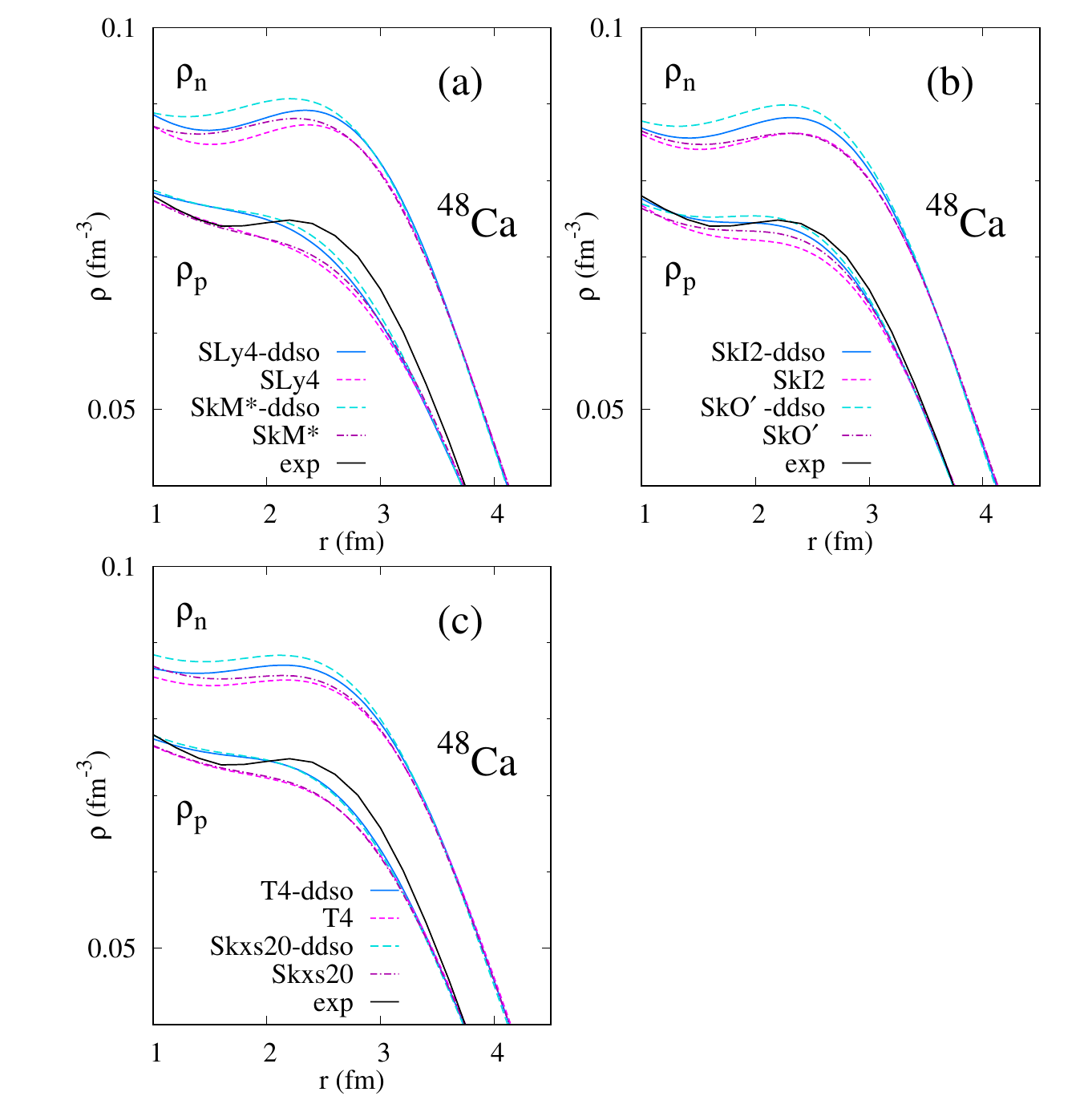}
\caption{
Neutron $(\rho_n)$ and proton $(\rho_p$)
densities in $\Ca48$ obtained from SHF calculations
using the Skyrme and Skyrme--ddso interactions,
together with the experimental proton density determined from electron-scattering data \cite{DeJager:1987qc}.
(The experimental values were read from figures in Refs.~\cite{Zenihiro:2018rmz,PhysRevC.102.064307}.)
\label{fig:dens-ddso-pn-ca48}}
\end{figure}

Figure~\ref{fig:pot-ddso-ca48} presents the Skyrme--ddso results for
the $\ell s$ potentials, single-particle densities, and neutron SPEs of $\Ca48$, and
Fig.~\ref{fig:dens-ddso-pn-ca48} compares the results for the neutron and proton densities 
with the original Skyrme results.
In the Skyrme--ddso results, the inner-surface neutron density around $r=2$--3~fm is increased
by the inclusion of the density-dependent spin--orbit term
because the
deeper $\ell s$ potential in this region increases the peak amplitude of the $1f_{7/2}$ neutron orbital.
The inner part of the surface proton density is also increased in the Skyrme--ddso results 
because the neutron density also contributes to the proton mean field. 
However, the proton density obtained depends on the Skyrme parameterizations, 
and agreement with the experimental data is not necessarily
satisfactory.

The inclusion of the density-dependent spin--orbit term somewhat affects the neutron SPEs
displayed in Fig.~\ref{fig:pot-ddso-ca48}(c).
Although the SPEs of the $1f_{7/2}$ and $2p_{3/2}$ orbitals are almost
unchanged, the $1f_{5/2}$ and $2p_{1/2}$ energies are decreased somewhat in the Skyrme--ddso results.
However, the change from the original Skyrme results is not as significant as it is for $\Pb208$.

\subsubsection{Differential mean-square charge radii of $\Ca48$ and $\Ca40$}

\begin{figure}[!htpb]
\includegraphics[width=8.6 cm]{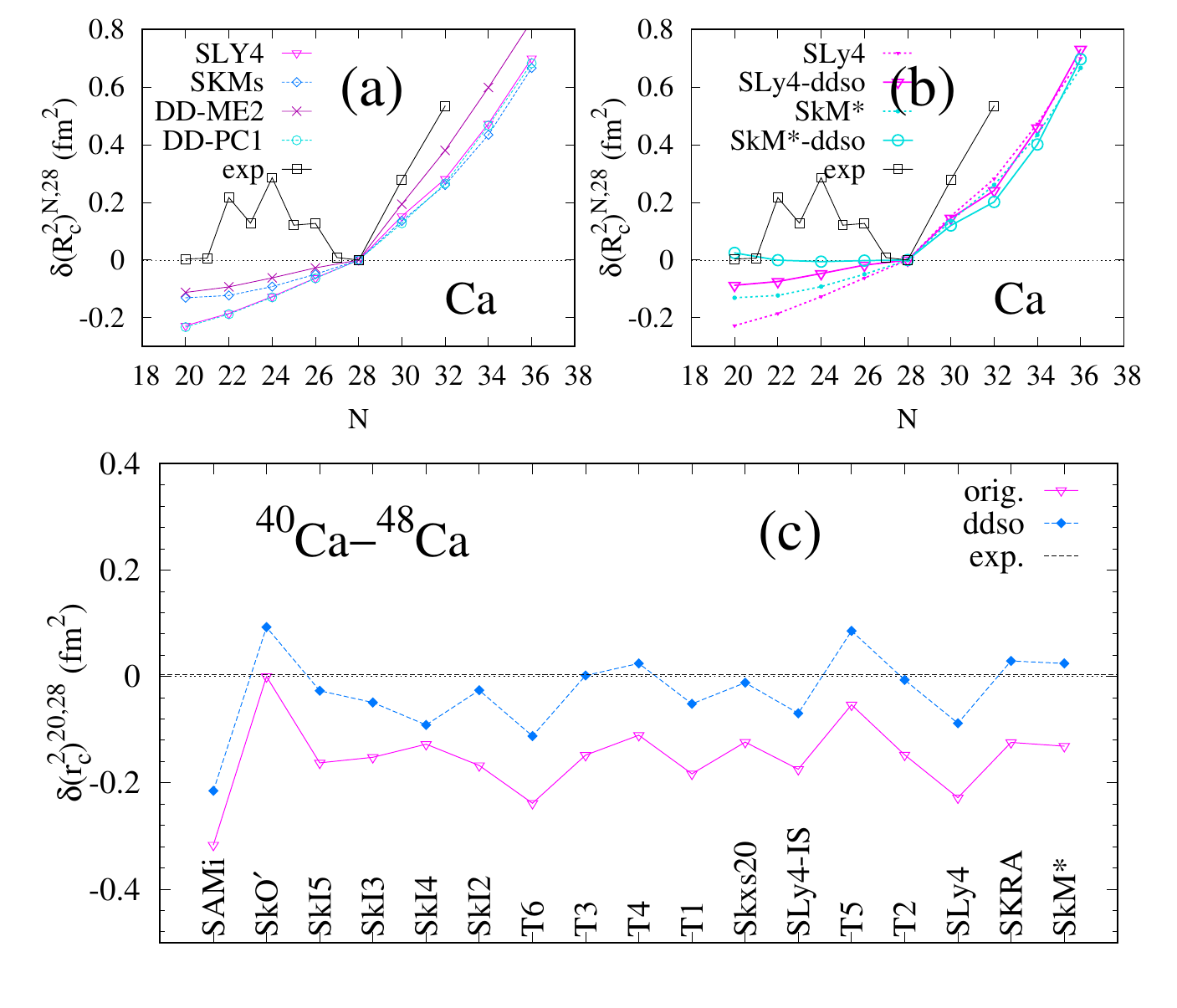}
\caption{(a) The differential mean-square charge radii $\delta(r^2_c)^\textrm{N,28}$ of the Ca isotopes
obtained from SHFB calculations using the SLy4 and SkM* interactions are
compared with the results obtained from RHB calculations using the DD--ME2 and DD--PC1 interactions.
(b) The SLy4--ddso and SkM*--ddso results together with the SLy4 and SkM* results.
(c) The values of $\delta(r^2_c)^\textrm{20,28}$ for the $\Ca40-\Ca48$ difference
obtained from SHF calculations using the Skyrme, Skyrme--ddso, and Skyrme--ddso2 interactions.
\label{fig:rc-ddso-caiso}}
\end{figure}

Figure~\ref{fig:rc-ddso-caiso}(a) presents
the differential mean-square charge radii
$\delta(r_c^2)^{N,28}$ of the Ca isotopes obtained from SHFB calculations
using the SLy4 and SkM* interactions, together with the DD--ME2 and DD--PC1 results from 
RHB calculations and the
experimental data. The experimental data for $\delta(r_c^2)^{N,28}$
show a kink behavior at $N=28$, meaning that the charge radius of $\Ca48$ is
abnormally small compared with those of the other Ca isotopes. Indeed,
$\delta(r_c^2)^{20,28}\sim 0$; i.e., the value of $r_c$ for $\Ca48$ is approximately
equal to that obtained for $\Ca40$.
The SHFB calculations using the SLy4 and SkM* interactions fail to
reproduce this kink behavior of $\delta(r_c^2)^{N,28}$. This is a known problem for
various versions of the Skyrme interactions.
The RHB calculations using the DD--ME2 and DD--PC1 interactions also fail to reproduce this behavior of $\delta(r_c^2)^{N,28}$.
The reason for the disagreement between the theoretical value of $\delta(r_c^2)^{N,28}$ and the data 
in the region $20<N<28$ is thought to be due to deformation and pairing effects.
However, the reason for the failure at $\delta(r_c^2)^{20,28}\sim 0$ for the doubly magic nucleus $\Ca48$ 
is not understood. 

To discuss the effects of the density-dependent spin--orbit term on the kink problem
of $\delta(r_c^2)^{N,28}$ at $N=28$ in Ca isotopes, I compare
the Skyrme and Skyrme--ddso results for $\delta(r_c^2)^{N,28}$
in Figs.~\ref{fig:rc-ddso-caiso}(b) and \ref{fig:rc-ddso-caiso}(c);
the former presents the values of $\delta(r_c^2)^{N,28}$ obtained
from SHFB calculations using the SLy4, SLy4--ddso, SkM*, and SkM*--ddso interactions, and
the latter presents the values of $\delta(r_c^2)^{20,28}$
obtained from SHF calculations using the Skyrme and Skyrme--ddso interactions, together with the
experimental data.
As Fig.~\ref{fig:rc-ddso-caiso}(c) shows,
the original Skyrme interactions generally underestimate the experimental value
$\delta(r_c^2)^{20,28}\sim 0$, except for the case of SkO$^\prime$.
In the Skyrme--ddso results, the values of $\delta(r_c^2)^{20,28}$ are increased
by the inclusion of the density-dependent spin--orbit term because of the contribution of the neutron $1f_{7/2}$ orbital to the proton mean field.

\section{Summary}\label{sec:summary}

I have investigated the $N$-dependence of the charge radii in the Pb isotopes and the inner part of the surface proton density of $\Pb208$
using SHFB and RHB calculations.
The conventional Skyrme interactions tend to
underestimate both the kink behavior of the charge radii at $N=126$ in the Pb isotopes
and the inner-surface proton density of $\Pb208$, whereas the RHB calculations
obtain better results.
I have shown that the kink behavior strongly is correlated with the energy difference
$e(1h_{11/2})-e(2g_{9/2})$ of the neutron $1h_{11/2}$ and $2g_{9/2}$ orbitals.
I performed detailed analyses of the SPEs and effective single-particle potentials obtained from the
SHFB and RHB calculations, and I found that the
essential difference between the non-relativistic and relativistic calculations involves the
$r$-dependence of the $\ell s$ potentials, which contribute significantly to
$e(1h_{11/2})-e(2g_{9/2})$.

To improve the SHFB calculations, I introduced a density-dependent spin--orbit term.
I obtained new Skyrme interactions that contain this term---called the
Skyrme--ddso interactions---by modifying the spin--orbit term in the
original Skyrme interactions. I compared the results obtained using the Skyrme--ddso interactions
with the original Skyrme results and have discussed the
effects of including this density-dependent spin--orbit term.
The results for the kink phenomena of the charge radii in the Pb isotopes were somewhat improved
by the inclusion of this term in the Skyrme--ddso calculations.
Moreover, the Skyrme--ddso calculations yield
better agreement with the experimental data for the proton density in the inner part of the surface region at $r\sim 5$~fm.
The change of the $\ell s$ potentials from the original Skyrme results
gives a significant contribution to the kink phenomenon
through its effects on the $\ell s$ splitting of the neutron $2g_{7/2,9/2}$ orbitals.
It also plays an important role in the inner parts of the surface proton densities through its effect on the
$1h_{11/2}$ proton orbitals.

I further discussed the contribution of the IS/IV ratio of the spin--orbit term.
In cases both with the density-dependent spin--orbit term---in
the Skyrme-ddso interactions---and without it---in the Skyrme interactions--- 
the change in the IS/IV ratio makes only minor contributions to the SPEs and to the kink phenomenon
in the Pb isotopes.

In addition, I investigated $\Ca48$ using SHF calculations with the Skyrme and Skyrme--ddso interactions
and discussed the effects of including the density-dependent spin--orbit term on the charge radii and densities.

In the present version of the Skyrme--ddso interactions, only the spin--orbit term was modified from the original Skyrme 
interactions, while the other parameters were left unchanged
in order to investigate the effects of the density-dependent spin--orbit term.
However, to construct a final version of new Skyrme interactions with a density-dependent spin--orbit term,
all the Skyrme parameters should be finely tuned by readjusting
the binding energies and radii of various nuclei over a wide region of the nuclear chart.

\begin{acknowledgments}
The author thanks Dr.~Hinohara for his help with the SHFB calculations.
Discussions during the YIPQS long-term international workshop
on ``Mean-field and Cluster Dynamics in Nuclear Systems" (MCD2022) were
useful in completing this work.
This work was supported
by Grants-in-Aid of the Japan Society
for the Promotion of Science (Grant Nos. JP18K03617,18H05407,	22K03633).
\end{acknowledgments}

\appendix

\section{Skyrme parameters}\label{sec:parameters}
The Skyrme energy densities are parametrized in terms of the quantities
$t_0$, $t_1$, $t_2$, $t_3$, $x_0$, $x_1$, $x_2$, $x_3$, $\gamma$, and $W_0$,
as shown in Eq.~(19) of Ref.~\cite{Bennaceur:2005mx}.
In some cases, the parameters $b_4$ and $b^\prime_4$ are used instead of $W_0$.
The set $b_4=b^\prime_4=W_0/2$ is equivalent to the usual parameterization of the spin--orbit term using $W_0$.
The values of these parameters for the Skyrme energy density
employed in the present paper are listed in Tables \ref{tab:skyrme-int1}
and \ref{tab:skyrme-int2}.
I employed the Skyrme parameters of the SLy4~\cite{Chabanat:1997un};
SkM*~\cite{Bartel:1982ed}; SkI2, SkI3, and SkI4~\cite{Reinhard:1995zz}; SAMi~\cite{Roca-Maza:2012dhp};
SkO$^\prime$~\cite{Reinhard:1999ut};
SKRA~\cite{Rashdan:2000}; Skxs20~\cite{Brown:2007zzc}; and SkT1 (T1), SkT2 (T2), SkT3 (T3), SkT4 (T4), SkT5 (T5), and SkT6 (T6)~\cite{Tondeur:1984gwk} interactions. In addition, I used modified versions of SLy4--- called the SLy4mod and SLy4--IS interactions---which contain an
IS-type spin--orbit term.
The former is taken from Ref.~\cite{Goddard:2012dk}, while the latter is
a new version obtained by adjusting the parameter $b_4$ to obtain the same spin--orbit energy
of $\Pb208$ as the original SLy4 result.

I adjusted 
selection parameters $\theta_\textrm{cm}$ and $\theta_{J^2}$ in the package of 
the HFBRAD code by choosing 
$\theta_\textrm{cm}=0$ and 1 for the calculations with and without the c.m. kinetic-energy correction,
and $\theta_{J^2}=0$ and 1 for the calculations without and with the $J^2$ term.
The SHFB calculations employed surface-type pairing forces.
For the SLy4 and SkM* interactions, I used the default parameterizations
in the HFBRAD code~\cite{Bennaceur:2005mx}.
For the other Skyrme interactions, the surface-type pairing forces of the SLy4 interaction in the HFBRAD code
were multiplied by a factor $\theta_\textrm{pairing}$, which I adjusted to give the 
mean neutron-pairing gap of $\Sn120$.
The values of $\theta_\textrm{cm}$, $\theta_{J^2}$, and $\theta_\textrm{pairing}$ are listed
in Tables \ref{tab:skyrme-int1} and \ref{tab:skyrme-int2}.
Note that---except for the SLy4 and SkM* interactions---these treatments are not necessarily the same as those in the original Skyrme calculations.

Tables \ref{tab:skyrme-int1} and \ref{tab:skyrme-int2} also list
the parameters of the Skyrme--ddso and Skyrme--ddso2 interactions:
$w_4$, $x_4$, and $\gamma_4$ for the density-dependent part of the spin--orbit term
and the multipliers $b_4/b_{4,\textrm{orig}}$ and $b^\prime_4/b_{4^\prime,\textrm{orig}}$
for the density-independent part.

\begin{table*}[!htpb]
\caption{
The Skyrme parameters $t_0$, $t_1$, $t_2$, $t_3$, $x_0$, $x1$, $x_2$, $x_3$, $\sigma$, $W_0$, $b_4$, and $b^\prime_4$ for the
SLy4~\cite{Chabanat:1997un};
SkM*~\cite{Bartel:1982ed}; SkI2, SkI3, and SkI4~\cite{Reinhard:1995zz}; SAMi~\cite{Roca-Maza:2012dhp};
SkO$^\prime$~\cite{Reinhard:1999ut};
SKRA~\cite{Rashdan:2000}; and Skxs20~\cite{Brown:2007zzc} interactions.
The selection parameters
$\theta_{J^2}$ and $\theta_\textrm{cm}$ in the HFBRAD code and the pairing factor $\theta_\textrm{pairing}$ adopted in the present paper
are also listed.
In addition, the values of $w_4$, $x_4$, and $\gamma_4$ for the density-dependent part of the spin--orbit term 
in the Skyrme--ddso and Skyrme--ddso2 interactions are listed.
For the Skyrme--ddso and Skyrme--ddso2 interactions,
the ratios of $b_4$ and $b^\prime_4$ to
the original values $b_{4,\textrm{orig.}}$ and $b^\prime_{4,\textrm{orig.}}$ in the Skyrme interactions for the
density-independent part of the spin--orbit term are also listed.
The units of the strength parameters $t_0$, $t_1,$ $t_2$, $t_3$, $W_0$, $b_4$, $b^\prime_4$, and $w_4$
are MeV, and the other parameters are dimensionless.
\label{tab:skyrme-int1}
}
\begin{center}
\begin{tabular}{rrrrrrrrrrrrrrrccccccccccc}
\hline
\hline
&	SLy4	&	SkM*	&	SkI2	&	SkI3	&	SkI4	&	SAMi	&	SkO'	&	SKRA	&	Skxs20	\\
$t_0$	&$	-2488.91	$&$	-2645	$&$	-1915.43	$&$	-1762.88	$&$	-1855.83	$&$	-1877.75	$&$	-2099.419	$&$	-2895.4	$&$	-2885.24	$\\
$t_1$	&$	486.82	$&$	410	$&$	438.449	$&$	561.608	$&$	473.829	$&$	475.6	$&$	301.531	$&$	405.5	$&$	302.73	$\\
$t_2$	&$	-546.39	$&$	-135	$&$	305.446	$&$	-227.09	$&$	1006.86	$&$	-85.2	$&$	154.781	$&$	-89.1	$&$	-323.42	$\\
$t_3$	&$	13777	$&$	15595	$&$	10548.9	$&$	8106.2	$&$	9703.61	$&$	10219.6	$&$	13526.464	$&$	16660	$&$	18237.49	$\\
$x_0$	&$	0.834	$&$	0.09	$&$	-0.2108	$&$	0.3083	$&$	0.4051	$&$	0.32	$&$	-0.029503	$&$	0.08	$&$	0.13746	$\\
$x_1$	&$	-0.344	$&$	0	$&$	-1.7376	$&$	-1.1722	$&$	-2.8891	$&$	-0.532	$&$	-1.325732	$&$	0	$&$	-0.25548	$\\
$x_2$	&$	-1	$&$	0	$&$	-1.5336	$&$	-1.0907	$&$	-1.3252	$&$	-0.014	$&$	-2.323439	$&$	0.2	$&$	-0.60744	$\\
$x_3$	&$	1.354	$&$	0	$&$	-0.178	$&$	1.2926	$&$	1.1452	$&$	0.688	$&$	-0.147404	$&$	0	$&$	0.05428	$\\
$\sigma$	&$	1/6	$&$	1/6	$&$	1/4	$&$	1/4	$&$	1/4	$&$	0.25614	$&$	1/4	$&$	0.1422	$&$	1/6	$\\
$W_0$	&$	123	$&$	130	$&$	120.602	$&$	0	$&$	0	$&$	0	$&$	0	$&$	129	$&$	0	$\\
$b_4$	&$ $&$ $&$ $&$	94.254	$&$	183.097	$&$	68.5	$&$	143.895	$&$ $&$	81.365	$\\
$b_4'$	&$ $&$ $&$ $&$	0	$&$	-180.351	$&$	21	$&$	-82.8888	$&$ $&$	0	$\\
$\theta_{J^2}$ &$	0	$&$	0	$&$	0	$&$	0	$&$	0	$&$	1	$&$	1	$&$	0	$&$	1	$\\
$\theta_\textrm{cm}$ &$	0	$&$	0	$&$	1	$&$	1	$&$	1	$&$	0	$&$	1	$&$	1	$&$	0	$\\
$\theta_\textrm{pairing}$	&	default	&	default	&	1	&	1.07	&	1.12	&	1.1	&	0.87	&	0.87	&	0.81	\\
& \\
\multicolumn{9}{c}{Skyrme-ddso}\\
$b_4/b_{4,\textrm{orig}}$ & 0.33 & 0.33 & 0.33 & 0.33 & 0.33 & 0.33 & 0.33 & 0.33 & 0.33\\
$b^\prime_4/b^\prime_{4,\textrm{orig}}$ & 0.33 & 0.33 & 0.33 & 0.33 & 0.33 & 0.33 & 0.33 & 0.33 & 0.33\\
$w_4$	&	840	&	870	&	835	&	865	&	815	&	745	&	915	&	860	&	740	\\
$x_4$	&	1	&	1	&	1	&	1	&	1	&	1	&	1	&	1	&	1	\\
$\gamma_4 $ &	1	&	1	&	1	&	1	&	1	&	1	&	1	&	1	&	1	\\
\multicolumn{9}{c}{Skyrme-ddso2}\\
$b_4/b_{4,\textrm{orig}}$ & 0.33 & 0.33 & 0.33 & 0.33 & 0.33 & 0.33 & 0.33 & 0.33 & 0.33\\
$b^\prime_4/b^\prime_{4,\textrm{orig}}$ & 0.33 & 0.33 & 0.33 & 0.33 & 0.33 & 0.33 & 0.33 & 0.33 & 0.33\\
$w_4$~(MeV)	&	1110	&	1145	&	1105	&	1145	&	1080	&	980	&	1210	&	1140	&	975	\\
$x_4$	&	0.5	&	0.5	&	0.5	&	0.5	&	0.5	&	0.5	&	0.5	&	0.5	&	0.5	\\
$\gamma_4 $ &	1	&	1	&	1	&	1	&	1	&	1	&	1	&	1	&	1	\\
\hline
\hline
\end{tabular}
\end{center}
\end{table*}

\begin{table*}[!htpb]
\caption{
The Skyrme parameters $t_0$, $t_1$, $t_2$, $t_3$, $x_0$, $x1$, $x_2$, $x_3$, $\sigma$, $W_0$, $b_4$, and $b^\prime_4$ for the
SkT1~(T1), SkT2~(T2), SkT3~(T3), SkT4~(T4), SkT5~(T5), and SkT6~(T6) ~\cite{Tondeur:1984gwk};
SLy4mod~\cite{Goddard:2012dk}; and SLy4--IS interactions.
The selection parameters
$\theta_{J^2}$ and $\theta_\textrm{cm}$ in the HFBRAD code and the pairing factor $\theta_\textrm{pairing}$ adopted in the present paper
are also listed.
In addition, the values of $w_4$, $x_4$, and $\gamma_4$ for the density-dependent part of the spin--orbit term
in the Skyrme--ddso and Skyrme--ddso2 interactions are listed.
For the Skyrme--ddso and Skyrme--ddso2 interactions,
the ratios of $b_4$ and $b^\prime_4$ to
the original values $b_{4,\textrm{orig.}}$ and $b^\prime_{4,\textrm{orig.}}$ of the Skyrme interactions for the
density-independent part of the spin--orbit term are also listed.
The units of the strength parameters $t_0$, $t_1,$ $t_2$, $t_3$, $W_0$, $b_4$, $b^\prime_4$, and $w_4$
are MeV, and the other parameters are dimensionless.
\label{tab:skyrme-int2}
}
\begin{center}
\begin{tabular}{rrrrrrrrrrrrrrrccccccccccc}
\hline
\hline
&	T1	&	T2	&	T3	&	T4	&	T5	&	T6	&	SLy4-IS	&	SLy4mod	\\
$t_0$	&$	-1794	$&$	-1791.6	$&$	-1791.8	$&$	-1808.8	$&$	-2917.1	$&$	-1794.2	$&$	-2488.91	$&$	-2488.91	$\\
$t_1$	&$	298	$&$	300	$&$	298.5	$&$	303.4	$&$	328.2	$&$	294	$&$	486.82	$&$	486.82	$\\
$t_2$	&$	-298	$&$	-300	$&$	-99.5	$&$	-303.4	$&$	-328.2	$&$	-294	$&$	-546.39	$&$	-546.39	$\\
$t_3$	&$	12812	$&$	12792	$&$	12794	$&$	12980	$&$	18584	$&$	12817	$&$	13777	$&$	13777	$\\
$x_0$	&$	0.154	$&$	0.154	$&$	0.138	$&$	-0.177	$&$	-0.295	$&$	0.392	$&$	0.834	$&$	0.834	$\\
$x_1$	&$	-0.5	$&$	-0.5	$&$	-1	$&$	-0.5	$&$	-0.5	$&$	-0.5	$&$	-0.344	$&$	-0.344	$\\
$x_2$	&$	-0.5	$&$	-0.5	$&$	1	$&$	-0.5	$&$	-0.5	$&$	-0.5	$&$	-1	$&$	-1	$\\
$x_3$	&$	0.089	$&$	0.089	$&$	0.075	$&$	-0.5	$&$	-0.5	$&$	0.5	$&$	1.354	$&$	1.354	$\\
$\sigma$ &$ 1/3 $&$ 1/3$&$ 1/3 $&$ 1/3 $&$ 1/6 $&$ 1/3 $&$ 1/6 $&$ $\\	&$	1/3	$&$	1/3	$&$	1/3	$&$	1/3	$&$	1/6	$&$	1/3	$&$	1/6	$&$	1/6	$\\
$W_0$	&$	110	$&$	120	$&$	126	$&$	113	$&$	114	$&$	107	$&$ $&$ $\\
$b_4$	&$ $&$ $&$ $&$ $&$ $&$ $&$	93	$&$	75	$\\
$b_4'$	&$ $&$ $&$ $&$ $&$ $&$ $&$	0	$&$	0	$\\
$\theta_{J^2}$	&$	1	$&$	1	$&$	1	$&$	1	$&$	1	$&$	1	$&$	0	$&$	0	$\\
$\theta_\textrm{cm}$	&$	0	$&$	0	$&$	0	$&$	0	$&$	0	$&$	0	$&$	0	$&$	0	$\\
$\theta_\textrm{pairing}$&	0.78	&	0.76	&	0.78	&	0.77	&	0.79	&	0.79	&	1.02	&	\\
& \\
\multicolumn{8}{c}{Skyrme-ddso}\\
$b_4/b_{4,\textrm{orig}}$ & 0.33 & 0.33 & 0.33 & 0.33 & 0.33 & 0.33 & 0.33 \\
$b^\prime_4/b^\prime_{4,\textrm{orig}}$ & 0.33 & 0.33 & 0.33 & 0.33 & 0.33 & 0.33 & 0.33 \\
$w_4$	&	750	&	805	&	855	&	785	&	790	&	725	&	840	& \\
$x_4$	&	1	&	1	&	1	&	1	&	1	&	1	&	1	& \\
$\gamma_4$ &	1	&	1	&	1	&	1	&	1	&	1	&	1	& \\
\multicolumn{9}{c}{Skyrme-ddso2}\\
$b_4/b_{4,\textrm{orig}}$ & 0.33 & 0.33 & 0.33 & 0.33 & 0.33 & 0.33 & 0.33 \\
$b^\prime_4/b^\prime_{4,\textrm{orig}}$ & 0.33 & 0.33 & 0.33 & 0.33 & 0.33 & 0.33 & 0.33 \\
$w_4$~(MeV)	&	995	&	1070	&	1130	&	1040	&	1045	&	960	&	1110	&	\\
$x_4$	&	0.5	&	0.5	&	0.5	&	0.5	&	0.5	&	0.5	&	0.5	& \\
$\gamma_4 $ &	1	&	1	&	1	&	1	&	1	&	1	&	1	& \\
\hline
\hline
\end{tabular}
\end{center}
\end{table*}

\bibliographystyle{apsrev4-1}
\bibliography{ddso-refs}

\end{document}